\documentclass[twocolumn,showpacs,amsmath,amssymb,10pt]{revtex4}
\usepackage{amsfonts}
\usepackage{bm}

\newcommand{\ot}{{\,\otimes\,}}
\newcommand{{\Cd}}{{\mathbb{C}^d}}

\def\oper{{\mathchoice{\rm 1\mskip-4mu l}{\rm 1\mskip-4mu l}%
{\rm 1\mskip-4.5mu l}{\rm 1\mskip-5mu l}}}
\def\<{\langle}
\def\>{\rangle}
\newtheorem{theorem}{Theorem}

\begin{document}
\title{\textbf{On circulant states with
positive partial transpose}} \author{Dariusz Chru\'sci\'nski and
Andrzej Kossakowski\thanks{email: darch@phys.uni.torun.pl} }
\affiliation{Institute of Physics, Nicolaus Copernicus University,\\
Grudzi\c{a}dzka 5/7, 87--100 Toru\'n, Poland}

\begin{abstract}
We construct a large  class of quantum $d \ot d$ states which are
positive under partial transposition (so called PPT states). The
construction is based on certain direct sum decomposition of the
total Hilbert space displaying characteristic circular structure
--- that is way we call them circulant states. It turns out that
partial transposition maps any such decomposition into another one
and hence both original density matrix and its partially
transposed partner share similar cyclic properties. This class
contains many well known examples of PPT states from the
literature and gives rise to a huge family of completely new
states.

\end{abstract}

\maketitle

\section{Introduction}

 The interest on  quantum entanglement has
dramatically increased during the last two decades due to the
emerging field of quantum information theory \cite{QIT}. It turns
out that quantum entanglement may be used as basic resources in
quantum information processing and communication. The prominent
examples are quantum cryptography, quantum teleportation, quantum
error correction codes and quantum computation.
\newline   \\
It is well known that it is extremely hard to check whether a
given density matrix describing a quantum state of the composite
system is separable or entangled.  There are several operational
criteria which enable one to detect quantum entanglement (see e.g.
\cite{Horodecki-review} for the recent review). The most famous
Peres-Horodecki criterion \cite{Peres,PPT} is based on the partial
transposition: if a state $\rho$ is separable then its partial
transposition $(\oper \ot \tau)\rho$ is positive. States which are
positive under partial transposition are called PPT states.
Clearly each separable state is necessarily PPT but the converse
is not true. It was shown by Horodecki et al. \cite{Horodeccy-PM}
that PPT condition is both necessary and sufficient for
separability for $2 \ot 2$ and $2 \ot 3$ systems.
\newline   \\
Now, since all separable states belong to a set of PPT states, the
structure of this set is of primary importance in quantum
information theory.  Unfortunately, this structure is still
unknown, that is,  one may check whether a given state is PPT but
we do not know how to construct a general quantum state with PPT
property.  There are only several  examples of PPT states which do
not show any systematic methods of constructing them (with one
exception, i.e. a class of PPT entangled states which is based on
a concept of unextendible product bases \cite{UPB} (see also
\cite{UPB-inne}). Other examples of PPT entangled states were
constructed in
\cite{PPT,Horodecki-book,Doherty-PPT,Shor,Ex2,Ex3,Ex4,Piani,Clarisse}
and the extreme points of the set of PPT states were recently
analyzed in \cite{Norwegia}. PPT states play also a crucial role
in mathematical theory of positive maps and, as is well know,
these maps are very important in the study of quantum
entanglement.  The mathematical structure of quantum entangled
states with positive partial transposition were studied in
\cite{Ha,Ha1,Ha2}.
\newline \\
Recently in \cite{III} we proposed a class of PPT states in $d \ot
d$ which are invariant under the maximal commutative subgroup of
$U(d)$, i.e. $d$-dimensional torus $U(1) \times \ldots \times
U(1)$. In the present paper we propose another class which defines
considerable generalization of \cite{III}. The construction of
this new class is based on certain  decomposition of the total
Hilbert space $\mathbb{C}^d \ot \mathbb{C}^d$ into direct sum of
$d$-dimensional subspaces. This decomposition is controlled by
some cyclic property, that is, knowing one subspace, say
$\Sigma_0$,  the  remaining subspaces $\Sigma_1, \ldots,
\Sigma_{d-1}$ are uniquely determined by applying a cyclic shift
to elements from $\Sigma_0$. Now, we call a density matrix $\rho$
a {\it circulant state} if $\rho$ is a convex combination of
density matrices supported on $\Sigma_\alpha$. The crucial
observation is that a partial transposition of the circulant state
has again a circular structure corresponding to another direct sum
decomposition $\widetilde{\Sigma}_0 \oplus \ldots \oplus
\widetilde{\Sigma}_{d-1}$.
\newline   \\
The paper is organized as follows: for pedagogical reason we first
illustrate our general method for $d=2$ in  Section \ref{2-QUBITS}
and for $d=3$ in Section \ref{2-QUTRITS}. Interestingly, there is
only one circular decomposition for $d=2$ and exactly two
different decompositions for $d=3$. In general case presented in
Section \ref{GENERAL} there are $(d-1)!$ decompositions labeled by
permutations from the symmetric group $S_{d-1}$. Section
\ref{EXAMPLES} presents several known examples of PPT states that
do belong to our class. Final conclusions are collected in the
last section.

\section{Two qubits} \label{2-QUBITS}

\subsection{An instructive example}

Consider a density matrix living in $\mathbb{C}^2 \ot
\mathbb{C}^2$ which has the following form:
\begin{equation}\label{2C}
    \rho = \left( \begin{array}{cc|cc}
    a_{00} & \cdot & \cdot & a_{01} \\
    \cdot      & b_{00} & b_{01} & \cdot \\ \hline
    \cdot      & b_{10} & b_{11} & \cdot \\
    a_{10} & \cdot & \cdot & a_{11} \end{array} \right)\ .
\end{equation}
In order to have more transparent pictures we replaced all
vanishing matrix elements by dots (we use this convention through
out this paper). It is clear that (\ref{2C}) defines a positive
operator iff the following $2\times 2$ matrices
\begin{equation}\label{}
    a = \left( \begin{array}{cc}
    a_{00} & a_{01} \\
    a_{10} & a_{11} \end{array} \right) \ , \ \ \ \
b = \left( \begin{array}{cc}
    b_{00} & b_{11} \\
    b_{10} & b_{11} \end{array} \right)\ ,
\end{equation}
are positive. Normalization adds additional condition
\[ {\rm Tr}\,a+ {\rm Tr}\,b = 1\ . \]
Now, the crucial observation is that partially transposed matrix
$\rho^\tau =(\oper \ot \tau)\rho$  belongs to the same class as
original $\rho$
\begin{equation}\label{2C-T}
    \rho^\tau = \left( \begin{array}{cc|cc}
    \widetilde{a}_{00} & \cdot & \cdot & \widetilde{a}_{01} \\
    \cdot      & \widetilde{b}_{00} & \widetilde{b}_{01} & \cdot \\ \hline
    \cdot      & \widetilde{b}_{10} & \widetilde{b}_{11} & \cdot \\
    \widetilde{a}_{10} & \cdot & \cdot & \widetilde{a}_{11} \end{array} \right)\
    ,
\end{equation}
where the matrices $\widetilde{a} = [\widetilde{a}_{ij}]$ and
$\widetilde{b} = [\widetilde{b}_{ij}]$ read as follows
\begin{equation}\label{}
    \widetilde{a} = \left( \begin{array}{cc}
    a_{00} & b_{01} \\
    b_{10} & a_{11} \end{array} \right) \ , \ \ \ \
\widetilde{b} = \left( \begin{array}{cc}
    b_{00} & a_{01} \\
    a_{10} & b_{11} \end{array} \right)\ .
\end{equation}
Hence, $\rho$ defined in (\ref{2C}) is PPT iff
\begin{equation*}\label{}
  \widetilde{a} \geq 0 \ \ \ \mbox{and} \ \ \ \widetilde{b} \geq 0
  \ .
\end{equation*}
The above conditions together with $a\geq 0$ and $b\geq 0$ may be
equivalently  rewritten as follows
\begin{eqnarray*}\label{}
a_{00}a_{11} & \geq & |a_{01}|^2\ , \\
a_{00}a_{11} & \geq & |b_{01}|^2\ ,
\end{eqnarray*}
and
\begin{eqnarray*}\label{}
b_{00}b_{11} & \geq & |a_{01}|^2\ , \\
b_{00}b_{00} & \geq & |b_{01}|^2\ ,
\end{eqnarray*}
which presents the full characterization of PPT states within a
class (\ref{2C}). We stress that for $b_{01}={b_{10}}=0$ the above
class reduces to the family of PPT states considered in
\cite{III}.

\subsection{Cyclic structure}

In order to generalize the above example to higher dimensional
cases let us observe that there is an interesting property of
cyclicity  which governs the structure of (\ref{2C}). For this
reason we call (\ref{2C}) circulant state. Note that $\rho$ may be
written as a direct sum
\begin{equation}\label{}
    \rho = \rho_0 +  \rho_1\ ,
\end{equation}
where $\rho_0$ and $\rho_1$ are supported on two orthogonal
subspaces
\begin{eqnarray}\label{}
    \Sigma_0 &=& \mbox{span}\left\{ e_0 \ot e_0\, , e_1 \ot e_1 \right\} \
    , \nonumber    \\
    \Sigma_1 &=& \mbox{span}\left\{ e_0 \ot e_1\, , e_1 \ot e_0 \right\} \ ,
\end{eqnarray}
where $\{e_0,e_1\}$ is a computational base in $\mathbb{C}^2$, and
clearly
\[  \Sigma_0 \oplus \Sigma_1\, = \, \mathbb{C}^2 \ot \mathbb{C}^2\ . \]
One has
\begin{eqnarray}\label{}
\rho_0 &=& \sum_{i,j=0}^1\, a_{ij}\, e_{ij} \ot e_{ij}\ , \\
\rho_1 &=& \sum_{i,j=0}^1\, b_{ij}\, e_{ij} \ot e_{i+1,j+1}\ ,
\end{eqnarray}
where
\begin{equation}\label{}
    e_{ij} = |e_i\>\<e_j|\ ,
\end{equation}
and one adds mod 2. Now, let us introduce the shift operator ${S}
: \mathbb{C}^2 \longrightarrow \mathbb{C}^2$ defined by
\begin{equation}\label{Shift-2}
    {S} \, e_i = e_{i+1}\ , \ \ \ \ \ (\mbox{mod}\ 2)\ .
\end{equation}
It is clear that matrix elements $S_{ij}$ define the following
circulant matrix \cite{C}
\begin{equation}\label{}
{S} = \left( \begin{array}{cc}
    \cdot & 1 \\
    1 & \cdot \end{array} \right) \ .
\end{equation}
One finds that
\begin{equation}\label{}
    \Sigma_1 = (\oper \ot S)\, \Sigma_0\ .
\end{equation}
Moreover, introducing two orthogonal projectors $P_0$ and $P_1=
(\oper \ot S) P_0(\oper \ot S)^*$ projecting onto $\Sigma_0$ and
$\Sigma_1$, respectively
\begin{eqnarray}\label{}
P_0 &=& \sum_{i=0}^1 e_{ii} \ot e_{ii} \  , \\
P_1 &=& \sum_{i=0}^1 e_{ii} \ot e_{i+1,i+1}\ ,
\end{eqnarray}
one finds
\begin{equation}\label{PPi}
\rho_i = P_i\,\rho\, P_i\ ,
\end{equation}
and hence
\begin{equation}\label{PPPPi}
    \rho = P_0\,\rho\, P_0 + P_1\,\rho\, P_1\ .
\end{equation}
Now, it turns out that (\ref{2C}) may be nicely rewritten in terms
of $S$. Introducing the following diagonal matrices
\begin{equation}\label{}
    x_{ij} = \left( \begin{array}{cc}
    a_{ij} & \cdot \\
    \cdot & b_{ij} \end{array} \right) \ ,
\end{equation}
one may rewrite (\ref{2C}) in the following form
\begin{equation}\label{2C-blocks}
    \rho = \left( \begin{array}{c|c}
    S^0\, x_{00}\, S^0 & S^0\, x_{01}\, S^1\\ \hline
    S^1\, x_{10}\, S^0 & S^1\, x_{11}\, S^1 \end{array} \right)\ .
\end{equation}
It is therefore clear  that partially transposed matrix
$\rho^\tau$ also possesses a cyclic structure
\begin{equation}\label{2C-blocks-T}
    \rho^\tau = \left( \begin{array}{c|c}
    S^0\, x_{00}\, S^0 & S^1\, x_{01}\, S^0\\ \hline
    S^0\, x_{10}\, S^1 & S^1\, x_{11}\, S^1 \end{array} \right)\ ,
\end{equation}
and may be decomposed as the following direct sum
\begin{equation}\label{}
    \rho^\tau = \widetilde{\rho}_0 + \widetilde{\rho}_1\ ,
\end{equation}
 with
\begin{eqnarray}\label{}
\widetilde{\rho}_0 &=& \sum_{i,j=0}^1\, \widetilde{a}_{ij}\, e_{ij} \ot e_{ij}\ , \\
\widetilde{\rho}_1 &=& \sum_{i,j=0}^1\, \widetilde{b}_{ij}\,
e_{ij} \ot e_{i+1,j+1}\ .
\end{eqnarray}
In analogy with (\ref{PPi}) and  (\ref{PPPPi}) one has
\begin{equation}\label{PPi}
\widetilde{\rho}_i = P_i\,\rho^\tau\, P_i\ ,
\end{equation}
and
\begin{equation}\label{PPPPi}
    \rho^\tau = P_0\,\rho^\tau\, P_0 + P_1\,\rho^\tau\, P_1\ .
\end{equation}
Note, that partial transposition $\rho \longrightarrow \rho^\tau$
reduces to the following operations on the level on $2\times 2$
matrices:
\[ a\ \longrightarrow\ \widetilde{a} \ \ \ \mbox{and} \ \ \ \ b
\longrightarrow\ \widetilde{b} \ . \] Again these operations are
fully controlled by the circulant matrix $S$
\begin{equation}\label{}
    \widetilde{a} = a \circ \mathbb{I} + b \circ S\ ,
\end{equation}
and similarly
\begin{equation}\label{}
    \widetilde{b} = b \circ \mathbb{I} + a \circ S\ ,
\end{equation}
where $x \circ y$ denotes the Hadamard product of two matrices $x$
and $y$ \cite{H}.

\section{Two qutrits}  \label{2-QUTRITS}

%\subsection{$3 \ot 3 = 3 \oplus 3 \oplus 3$}

A similar construction may be performed in $\mathbb{C}^3 \ot
\mathbb{C}^3$. The basic idea is to decompose the total Hilbert
space $\mathbb{C}^3 \ot \mathbb{C}^3$ into a direct sum of three
orthogonal subspaces $\Sigma_i$ related by a certain cyclic
property. In analogy to (\ref{Shift-2}) let us define a shift
operator ${S} : \mathbb{C}^3 \longrightarrow \mathbb{C}^3$ via
\begin{equation}\label{Shift-3}
    {S} \, e_i = e_{i+1}\ , \ \ \ \ \ (\mbox{mod}\ 3)\ .
\end{equation}
It is clear that matrix elements $S_{ij}$ define the following $3
\times 3$ circulant matrix
\begin{equation}\label{}
{S} = \left( \begin{array}{ccc}
    \cdot &  \cdot & 1 \\
    1 & \cdot & \cdot  \\
     \cdot & 1&  \cdot \end{array} \right) \ .
\end{equation}
Now, let us define three orthogonal 3-dimensional subspaces in
$\mathbb{C}^3 \ot \mathbb{C}^3$
\begin{equation}\label{}
     \Sigma_0 = \mbox{span}\left\{ e_0 \ot e_0\, , e_1 \ot e_1\, , e_2 \ot e_2 \right\} \ ,
\end{equation}
and
\begin{equation}\label{}
    \Sigma_1 = (\oper \ot S)\Sigma_0\ , \ \ \ \
\Sigma_2 = (\oper \ot S^2)\Sigma_0\ .
\end{equation}
One easily finds
\begin{eqnarray}\label{}
    \Sigma_1 &=& \mbox{span}\left\{ e_0 \ot e_1\, , e_1 \ot e_2\, , e_2\ot e_0 \right\} \
    ,\nonumber  \\
\Sigma_2 &=& \mbox{span}\left\{ e_0 \ot e_2\, , e_1 \ot e_0\, ,
e_2\ot e_1 \right\} \ ,
\end{eqnarray}
together with
\[  \Sigma_0 \oplus \Sigma_1 \oplus \Sigma_2\, = \, \mathbb{C}^3 \ot \mathbb{C}^3\ . \]
The construction of a circulant state in $\mathbb{C}^3 \ot
\mathbb{C}^3$ goes as follows: define three positive operators
$\rho_i$ which are supported on $\Sigma_i$ $(i=0,1,2)$:
\begin{eqnarray}\label{}
\rho_0 &=& \sum_{i,j=0}^2\, a_{ij}\, e_{ij} \ot e_{ij}\ , \\
\rho_1 &=& \sum_{i,j=0}^2\, b_{ij}\, e_{ij} \ot S\,e_{ij}\, S^*\ ,\nonumber\\
       &=& \sum_{i,j=0}^2\, b_{ij}\, e_{ij} \ot e_{i+1,j+1}\, \\
\rho_2 &=& \sum_{i,j=0}^2\, c_{ij}\, e_{ij} \ot S^2\,e_{ij}\, S^{*2}\ , \nonumber\\
        &=&\sum_{i,j=0}^2\, c_{ij}\, e_{ij} \ot e_{i+2,j+2}\ ,
\end{eqnarray}
where $a_{ij}$, $b_{ij}$ and $c_{ij}$ give rise to the following
$3 \times 3$ matrices:
\begin{eqnarray*}\label{}
a\  =\  \left( \begin{array}{ccc}
    a_{00} & a_{01} & a_{02} \\
    a_{10} & a_{11} & a_{12} \\
    a_{20} & a_{21} & a_{22} \end{array} \right) \ , \ \ \ \
b \ =\ \left( \begin{array}{ccc}
    b_{00} & b_{01} & b_{02} \\
    b_{10} & b_{11} & b_{12} \\
    b_{20} & b_{21} & b_{22} \end{array} \right)\ ,
\end{eqnarray*}
\begin{equation*}\label{}
c\ =\ \left( \begin{array}{ccc}
    c_{00} & c_{01} & c_{02} \\
    c_{10} & c_{11} & c_{12} \\
    c_{20} & c_{21} & c_{22} \end{array} \right)\ .
\end{equation*}
Positivity of $\rho_i$ is guaranteed by positivity of $a$, $b$ and
$c$. Finally, define a circulant $3 \ot 3$ state by
\begin{equation}\label{}
    \rho = \rho_0 +  \rho_1 + \rho_2\ .
\end{equation}
It is clear that
\[   \rho_\alpha = P_\alpha\, \rho \, P_\alpha\ , \ \ \ \ \
\alpha=0,1,2\ , \]
where $P_\alpha$ denotes orthogonal projector
onto $\Sigma_\alpha$:
\begin{equation*}\label{}
    P_0 = \sum_{i=0}^2 e_{ii} \ot e_{ii} \ ,
\end{equation*}
and
\[ P_\alpha = (\oper \ot S^\alpha)\, P_0 \, (\oper \ot
S^*)^\alpha\ , \] with
\[ P_0 + P_1 + P_2 = \mathbb{I} \ot \mathbb{I}\ . \]
 Using definitions of $\rho_i$ one easily finds
\begin{equation}\label{3C}
 \hspace*{-.1cm}
  \rho = \left( \begin{array}{ccc|ccc|ccc}
    a_{00} & \cdot & \cdot & \cdot & a_{01} & \cdot & \cdot & \cdot & a_{02} \\
    \cdot& b_{00} & \cdot & \cdot & \cdot& b_{01} & b_{02} & \cdot & \cdot  \\
    \cdot& \cdot& c_{00} & c_{01} & \cdot & \cdot & \cdot & c_{02} &\cdot   \\ \hline
    \cdot & \cdot & c_{10} & c_{11} & \cdot & \cdot & \cdot & c_{12} & \cdot \\
    a_{10} & \cdot & \cdot & \cdot & a_{11} & \cdot & \cdot & \cdot & a_{12}  \\
    \cdot& b_{10} & \cdot & \cdot & \cdot & b_{11} & b_{12} & \cdot & \cdot  \\ \hline
    \cdot & b_{20} & \cdot & \cdot& \cdot & b_{21} & b_{22} & \cdot & \cdot \\
    \cdot& \cdot & c_{20} & c_{21} & \cdot& \cdot & \cdot & c_{22} & \cdot  \\
    a_{20} & \cdot& \cdot & \cdot & a_{21} & \cdot& \cdot & \cdot & a_{22}
     \end{array} \right)\ .
\end{equation}
Normalization of $\rho$ implies
\[  \mbox{Tr}\Big( a + b + c \Big) = 1\ . \]
 It turns out that (\ref{3C}) may be nicely rewritten in terms
of $S$. Introducing the following diagonal matrices
\begin{equation}\label{}
    x_{ij} = \left( \begin{array}{ccc}
    a_{ij} & \cdot & \cdot \\
    \cdot & b_{ij} & \cdot\\
    \cdot & \cdot & c_{ij} \end{array} \right) \ ,
\end{equation}
one may rewrite (\ref{3C}) in the following block form
\begin{equation}\label{3C-blocks}
    \rho = \left( \begin{array}{c|c|c}
    S^0\, x_{00}\, S^{*0} & S^0\, x_{01}\, S^{*1} & S^0\, x_{02}\, S^{*2} \\ \hline
    S^1\, x_{10}\, S^{*0} & S^1\, x_{11}\, S^{*1} & S^1\, x_{12}\, S^{*2} \\ \hline
    S^2\, x_{20}\, S^{*0} & S^2\, x_{21}\, S^{*1} & S^2\, x_{22}\, S^{*2}\end{array} \right)\ .
\end{equation}
Partially transposed $\rho^\tau$ has the following form
\begin{equation}\label{3C-T}
 \hspace*{-.1cm}
  \rho^\tau = \left( \begin{array}{ccc|ccc|ccc}
    \widetilde{a}_{00} & \cdot & \cdot & \cdot &  \cdot & \widetilde{a}_{01} & \cdot &  \widetilde{a}_{02} & \cdot  \\
    \cdot& \widetilde{b}_{00} & \cdot & \widetilde{b}_{01} & \cdot & \cdot&   \cdot & \cdot & \widetilde{b}_{02}  \\
    \cdot& \cdot& \widetilde{c}_{00} &  \cdot & \widetilde{c}_{01} & \cdot & \widetilde{c}_{02} & \cdot & \cdot   \\ \hline
    \cdot & \widetilde{b}_{10} & \cdot &  \widetilde{b}_{11} & \cdot & \cdot & \cdot  & \cdot & \widetilde{b}_{12} \\
     \cdot & \cdot & \widetilde{c}_{10} & \cdot & \widetilde{c}_{11} & \cdot & \widetilde{c}_{12} & \cdot & \cdot   \\
    \widetilde{a}_{10} & \cdot & \cdot & \cdot & \cdot & \widetilde{a}_{11} & \cdot & \widetilde{a}_{12} & \cdot  \\ \hline
    \cdot &  \cdot & \widetilde{c}_{20} & \cdot & \widetilde{c}_{21} & \cdot &  \widetilde{c}_{22} & \cdot & \cdot \\
    \widetilde{a}_{20} & \cdot & \cdot & \cdot & \cdot & \widetilde{a}_{21} & \cdot & \widetilde{a}_{22} & \cdot  \\
     \cdot & \widetilde{b}_{20} & \cdot & \widetilde{b}_{21} & \cdot &  \cdot& \cdot & \cdot & \widetilde{b}_{22}
     \end{array} \right)\ ,
\end{equation}
where the matrices $\widetilde{a} = [\widetilde{a}_{ij}]$,
$\widetilde{b} = [\widetilde{b}_{ij}]$ and $\widetilde{c} =
[\widetilde{c}_{ij}]$ read as follows

\begin{eqnarray*}\label{}
\widetilde{a} &=& \left( \begin{array}{ccc}
    a_{00} & c_{01} & b_{02} \\
    c_{10} & b_{11} & a_{12} \\
    b_{20} & a_{21} & c_{22} \end{array} \right) \ , \ \ \
\widetilde{b} \ = \left( \begin{array}{ccc}
    b_{00} & a_{01} & c_{02} \\
    a_{10} & c_{11} & b_{12} \\
    c_{20} & b_{21} & a_{22} \end{array} \right)\ ,
\end{eqnarray*}
\begin{equation*}\label{}
\widetilde{c} \ = \ \left( \begin{array}{ccc}
    c_{00} & b_{01} & a_{02} \\
    b_{10} & a_{11} & c_{12} \\
    a_{20} & c_{21} & b_{22} \end{array} \right)\ .
\end{equation*}
Note, that
\begin{equation}\label{}
    \rho^\tau = \widetilde{\rho}_0 +  \widetilde{\rho}_1 + \widetilde{\rho}_2\
    ,
\end{equation}
where $\widetilde{\rho}_k$ are supported on three orthogonal
subspaces of $\mathbb{C}^3 \ot \mathbb{C}^3$:
\begin{eqnarray}\label{}
 \widetilde{\Sigma}_0 &=& \mbox{span}\left\{ e_0 \ot e_0\, , e_1 \ot e_2\, , e_2 \ot e_1 \right\} \
 ,\nonumber\\
    \widetilde{\Sigma}_1 &=& \mbox{span}\left\{ e_0 \ot e_1\, , e_1 \ot e_0\, , e_2\ot e_2 \right\} \
    ,\\
\widetilde{\Sigma}_2 &=& \mbox{span}\left\{ e_0 \ot e_2\, , e_1
\ot e_1\, , e_2\ot e_0 \right\} \ .\nonumber
\end{eqnarray}
One has therefore
\begin{theorem}
A circulant $3 \ot 3$ state $\rho$ is {\rm PPT} iff the matrices
$\widetilde{a}$, $\widetilde{b}$ and $\widetilde{c}$ are positive.
\end{theorem}

 Note, that
\[ \widetilde{\Sigma}_0 \ = \ (\oper \ot \widetilde{\Pi})\, \Sigma_0\ , \]
where $\widetilde{\Pi}$ is the following permutation matrix
\begin{equation}\label{}
    \widetilde{\Pi}\ = \ \left( \begin{array}{ccc}
    1 & \cdot &  \cdot  \\
     \cdot & \cdot & 1  \\
     \cdot & 1 &  \cdot \end{array} \right) \ .
\end{equation}
 Again, one has a cyclic structure
\begin{equation}\label{}
\widetilde{\Sigma}_i = (\oper \ot S) \widetilde{\Sigma}_0\ .
\end{equation}
Moreover, it is clear that
\[   \widetilde{\rho}_\alpha = \widetilde{P}_\alpha\, \rho^\tau \, \widetilde{P}_\alpha\ , \ \ \ \ \
\alpha=0,1,2\ , \] where $\widetilde{P}_\alpha$ denotes orthogonal
projector onto $\widetilde{\Sigma}_\alpha$:
\begin{equation*}\label{}
    \widetilde{P}_0 = (\oper \ot \Pi) \, P_0 \, (\oper \ot
\Pi^*)\ ,
\end{equation*}
and
\[ \widetilde{P}_\alpha = (\oper \ot S^\alpha)\, \widetilde{P}_0 \, (\oper \ot
S^*)^\alpha\ , \] with
\[ \widetilde{P}_0 + \widetilde{P}_1 + \widetilde{P}_2 = \mathbb{I} \ot \mathbb{I}\ . \]
 It is therefore clear that $\rho^\tau$ is again
a circulant operator and its circular structure is governed by
\begin{equation}\label{3C-blocks-T}
    \rho^\tau = \left( \begin{array}{c|c|c}
    {S}^0\, \widetilde{x}_{00}\, {S}^{*0}
    & {S}^0\, \widetilde{x}_{01}\, {S}^{*2}
    & {S}^0\, \widetilde{x}_{02}\, {S}^{*1} \\ \hline
    {S}^2\, \widetilde{x}_{10}\, {S}^{*0}
    & {S}^2\, \widetilde{x}_{11}\, {S}^{*2}
    & {S}^2\, \widetilde{x}_{12}\, {S}^{*1} \\ \hline
    {S}^1\, \widetilde{x}_{20}\, {S}^{*0}
    & {S}^1\, \widetilde{x}_{21}\, {S}^{*2}
    & {S}^1\, \widetilde{x}_{22}\, {S}^{*1}\end{array} \right)\
    ,
\end{equation}
where
\begin{equation}\label{}
    \widetilde{x}_{ij} = \left( \begin{array}{ccc}
    \widetilde{a}_{ij} & \cdot & \cdot \\
    \cdot & \widetilde{b}_{ij} & \cdot\\
    \cdot & \cdot & \widetilde{c}_{ij} \end{array} \right) \ .
\end{equation}
Interestingly, matrices $\widetilde{a}$, $\widetilde{b}$ and
$\widetilde{c}$ may be nicely defined in terms of
$\widetilde{\Pi}$ and ${S}$. It is not difficult to show that
\begin{eqnarray}\label{}
\widetilde{a} & =& a \circ \widetilde{\Pi} + b \circ
(\widetilde{\Pi}\,{S}) + c \circ
(\widetilde{\Pi}\,{S}^2)\ , \nonumber \\
\widetilde{b} & =& b \circ \widetilde{\Pi} + c \circ
(\widetilde{\Pi}\,{S}) + a \circ
(\widetilde{\Pi}\,{S}^2)\ ,  \\
\widetilde{c} & =& c \circ \widetilde{\Pi} + a \circ
(\widetilde{\Pi}\,{S}) + b \circ (\widetilde{\Pi}\,{S}^2)\ ,
\nonumber
\end{eqnarray}
where ``$\circ$" denotes the Hadamard product.

Let us stress that this class in a significant way enlarges the
class considered in \cite{III}. One reconstruct \cite{III} by
taking as $b$ and $c$ diagonal matrices:
\begin{eqnarray*}\label{}
b\  =\  \left( \begin{array}{ccc}
    b_{00} & \cdot & \cdot \\
    \cdot & b_{11} & \cdot \\
    \cdot & \cdot & b_{22} \end{array} \right) \ , \ \ \ \
c \ =\ \left( \begin{array}{ccc}
     c_{00} & \cdot & \cdot \\
    \cdot & c_{11} & \cdot \\
    \cdot & \cdot & c_{22} \end{array} \right)\ .
\end{eqnarray*}
Then one finds for partially transposed matrix:
\begin{eqnarray*}\label{}
\widetilde{a} &=& \left( \begin{array}{ccc}
    a_{00} & \cdot & \cdot \\
    \cdot & b_{11} & a_{12} \\
    \cdot & a_{21} & c_{22} \end{array} \right) \ , \ \ \
\widetilde{b} \ = \left( \begin{array}{ccc}
    b_{00} & a_{01} & \cdot \\
    a_{10} & c_{11} & \cdot \\
    \cdot & \cdot & a_{22} \end{array} \right)\ ,
\end{eqnarray*}
and
\begin{equation*}\label{}
\widetilde{c} \ = \ \left( \begin{array}{ccc}
    c_{00} & \cdot & a_{02} \\
    \cdot & a_{11} & \cdot \\
    a_{20} & \cdot & b_{22} \end{array} \right)\ .
\end{equation*}

\section{General $\ d\ot d\ $ case}   \label{GENERAL}

Now we are ready to construct circular states in $d \ot d$. The
basic idea is to decompose the total Hilbert space $\mathbb{C}^d
\ot \mathbb{C}^d$ into a direct sum of $d$ orthogonal
$d$-dimensional subspaces related by a certain cyclic property. It
turns out that there are $(d-1)!$ different cyclic decompositions
and it is therefore clear that they may be labeled by permutations
from the symmetric group $S_{d-1}$. For $d=2$ one has only one
decomposition
\begin{eqnarray}\label{Dec-d=2}
    \Sigma_0 &=& \mbox{span}\left\{ e_0 \ot e_0\, , e_1 \ot e_1 \right\} \
    , \nonumber    \\
    \Sigma_1 &=& \mbox{span}\left\{ e_0 \ot e_1\, , e_1 \ot e_0 \right\} \ ,
\end{eqnarray}
whereas for $d=3$ we have found 2 different cyclic decompositions
\begin{eqnarray}\label{Dec-d=3|1}
\Sigma_0 &=& \mbox{span}\left\{ e_0 \ot e_0\, , e_1 \ot e_1\, ,
e_2 \ot e_2 \right\} \ ,    \nonumber \\
    \Sigma_1 &=& \mbox{span}\left\{ e_0 \ot e_1\, , e_1 \ot e_2\, , e_2\ot e_0 \right\} \
    ,\\
\Sigma_2 &=& \mbox{span}\left\{ e_0 \ot e_2\, , e_1 \ot e_0\, ,
e_2\ot e_1 \right\} \ , \nonumber
\end{eqnarray}
and
\begin{eqnarray}\label{Dec-d=3|2}
 \widetilde{\Sigma}_0 &=& \mbox{span}\left\{ e_0 \ot e_0\, , e_1 \ot e_2\, , e_2 \ot e_1 \right\} \
\nonumber  ,\\
    \widetilde{\Sigma}_1 &=& \mbox{span}\left\{ e_0 \ot e_1\, , e_1 \ot e_0\, , e_2\ot e_2 \right\} \
    ,\\
\widetilde{\Sigma}_2 &=& \mbox{span}\left\{ e_0 \ot e_2\, , e_1
\ot e_1\, , e_2\ot e_0 \right\} \ . \nonumber
\end{eqnarray}
Let us introduce a basic $d$-dimensional subspace
\begin{equation}\label{}
{\Sigma}_0 = \mbox{span}\left\{ e_0 \ot e_0\, , e_1 \ot e_1\, ,
\ldots\, , e_{d-1} \ot e_{d-1} \right\} \ .
\end{equation}
Now, for any permutation $\pi \in S$ let us define
${\Sigma}_0^\pi$ which is spanned by
\begin{equation}\label{}
 e_0 \ot e_{\pi(0)}\, , e_1 \ot e_{\pi(1)}\, , \ldots\, , e_{d-1}
\ot e_{\pi(d-1)} \ .
\end{equation}
Note, that introducing a permutation matrix $\Pi$ corresponding to
$\pi$ one has
\begin{equation}\label{}
{\Sigma}_0^\pi = (\oper \ot \Pi) \Sigma_0\ .
\end{equation}
Actually, it is enough  to consider only a subset of permutations
such that $\pi(0)=0$, it means that vector $e_0 \ot e_0$ always
belongs to the subspace number `0' in each decomposition. Finally,
the remaining $(d-1)$ subspaces in the decomposition labeled by
$\pi$ are  defined via
\begin{eqnarray}\label{}
{\Sigma}^\pi_\alpha &=& (\oper \ot S^\alpha) {\Sigma}^\pi_0\ ,
\nonumber \\
&=& (\oper \ot S^\alpha\, \Pi) {\Sigma}_0\ ,
\end{eqnarray}
where $S$ is a circulant matrix corresponding to shift in
$\mathbb{C}^d$:
\begin{equation}\label{}
    S = \left( \begin{array}{ccccc}
    \cdot & \cdot & \ldots & \cdot & 1 \\
    1 &    \cdot & \ldots & \cdot & \cdot \\
    \cdot & 1 & \ldots & \cdot & \cdot \\
    \vdots & \vdots& \ddots & \vdots & \vdots \\
    \cdot & \cdot & \ldots & 1 & \cdot \end{array} \right)\  .
\end{equation}
One easily check
\[ {\Sigma}^\pi_0 \oplus {\Sigma}^\pi_1 \oplus \ldots \oplus
{\Sigma}^\pi_{d-1}\, =\, \mathbb{C}^d \ot \mathbb{C}^d\ . \]

To construct a circulant state corresponding to this decomposition
let us introduce $d$ positive $d \times d$ matrices $a^{(\alpha)}
= [ a^{(\alpha)}_{ij}]\, ; \, \alpha=0,1,\ldots,d-1$. Now, define
$d$ positive operators $\rho^\pi_\alpha$ supported on
${\Sigma}^\pi_\alpha$ via
\begin{eqnarray}\label{}
\rho^\pi_\alpha &=& \sum_{i,j=0}^{d-1}\, a^{(\alpha)}_{ij}\,
e_{ij} \ot S^\alpha\, e_{\pi(i),\pi(j)}\, S^{*\alpha} \nonumber \\
&=& \sum_{i,j=0}^{d-1}\, a^{(\alpha)}_{ij}\, e_{ij} \ot
e_{\pi(i)+\alpha,\pi(j)+\alpha}\ .
\end{eqnarray}
Finally,
\begin{equation}\label{}
    \rho_\pi = \rho^\pi_0 + \rho^\pi_1 + \ldots +
    \rho^\pi_{d-1}\ ,
\end{equation}
defines circulant state (corresponding to $\pi$). Normalization of
$\rho_\pi$ is equivalent to the following condition for matrices
$a^{(\alpha)}$
\[   \mbox{Tr}\, \left( a^{(0)} + a^{(1)} + \ldots +
a^{(d-1)} \right) = 1\ . \]

Interestingly, $\rho_\pi$ has the following transparent block
form: introduce a set of $d^2$ diagonal matrices
\begin{equation}\label{}
x_{ij} = \left( \begin{array}{cccc}
    a^{(0)}_{ij} & \cdot & \ldots & \cdot  \\
    \cdot & a^{(1)}_{ij} & \ldots & \cdot  \\
    \vdots & \vdots & \ddots & \vdots  \\
    \cdot & \cdot & \ldots & a^{(d-1)}_{ij} \end{array} \right)\  ,
\end{equation}
then one finds
\begin{widetext}
\begin{equation}\label{rho-d-block}
\rho_\pi\, =\, \left( \begin{array}{c|c|c|c}
    S^{\pi(0)} x_{00} S^{\pi(0)*} & S^{\pi(0)} x_{01} S^{\pi(1)*} &\ \ \
    \ldots\ \ \  & S^{\pi(0)} x_{0,d-1} S^{\pi(d-1)*}  \\ \hline
    S^{\pi(1)} x_{10} S^{\pi(0)*} & S^{\pi(1)} x_{11} S^{\pi(1)*} &
    \ldots & S^{\pi(1)} x_{1,d-1} S^{\pi(d-1)*}    \\ \hline
    \vdots & \vdots & \ddots & \vdots  \\ \hline
    S^{\pi(d-1)} x_{d-1,0} S^{\pi(0)*} & S^{\pi(d-1)} x_{d-1,1} S^{\pi(1)*} &
    \ldots & S^{\pi(d-1)} x_{d-1,d-1} S^{\pi(d-1)*}
         \end{array} \right)\  .
\end{equation}
\end{widetext}
 Having defined a circulant state $\rho_\pi$ let us look for
a partially transposed matrix $\rho_\pi^\tau$. Now comes the
crucial observation

\begin{theorem}
If $\rho_\pi$ is a circulant state corresponding to permutation
$\pi$ such that $\pi(0)=0$, then its partial transposition
$\rho_\pi^\tau$ is also circulant with respect to another
decomposition corresponding to permutation $\widetilde{\pi}$ such
that
\begin{equation}\label{pi-pi}
    \pi(i) + \widetilde{\pi}(i) = d\ ,
\end{equation}
for $i=1,2,\ldots,d-1$, and $\widetilde{\pi}(0)=0$.
\end{theorem}

Note, that for $d=2$ there is only one (trivial) permutation
$(\pi(0)=0,\pi(1)=1)$ and hence $\widetilde{\pi}=\pi$, that is
both $\pi$ and $\widetilde{\pi}$ define the same decomposition
(\ref{Dec-d=2}). For $d=3$ one has two different permutations in
$S_2$: the trivial one $(\pi(0)=0,\pi(1)=1,\pi(2)=2)$ which
corresponds to (\ref{Dec-d=3|1}) and "true" permutation
$(\widetilde{\pi}(0)=0,\widetilde{\pi}(1)=2,\widetilde{\pi}(2)=1)$
which corresponds to (\ref{Dec-d=3|2}).

Hence, $\rho_\pi^\tau$ may be decomposed as follows
\begin{equation}\label{}
    \rho_\pi^\tau = \widetilde{\rho}^\pi_0 + \widetilde{\rho}^\pi_1 + \ldots +
    \widetilde{\rho}^\pi_{d-1}\ ,
\end{equation}
and $\widetilde{\rho}^\pi_\alpha$ are defined by
\begin{eqnarray}\label{}
\widetilde{\rho}^\pi_\alpha &=& \sum_{i,j=0}^{d-1}\,
\widetilde{a}^{(\alpha)}_{ij}\,
e_{ij} \ot S^\alpha\, e_{\widetilde{\pi}(i),\widetilde{\pi}(j)}\, S^{*\alpha} \nonumber \\
&=& \sum_{i,j=0}^{d-1}\, \widetilde{a}^{(\alpha)}_{ij}\, e_{ij}
\ot e_{\widetilde{\pi}(i)+\alpha,\widetilde{\pi}(j)+\alpha}\ ,
\end{eqnarray}
where again we trivially extended $\widetilde{\pi}$ from $S_{d-1}$
to $S_d$ by  $\widetilde{\pi}(0)\equiv 0$.

In analogy to (\ref{rho-d-block}) one finds the following block
form of $\rho_\pi^\tau$:
\begin{widetext}
\begin{equation}\label{rho-d-block-T}
\rho^\tau_\pi\, =\, \left( \begin{array}{c|c|c|c}
S^{\widetilde{\pi}(0)} x_{00} S^{\widetilde{\pi}(0)*} &
S^{\widetilde{\pi}(0)} x_{01} S^{\widetilde{\pi}(1)*} &\ \ \
    \ldots\ \ \  & S^{\widetilde{\pi}(0)} x_{0,d-1} S^{\widetilde{\pi}(d-1)*}_\pi  \\ \hline
    S^{\widetilde{\pi}(1)} x_{10} S^{\widetilde{\pi}(0)*} & S^{\widetilde{\pi}(1)} x_{11} S^{\widetilde{\pi}(1)*} &
        \ldots & S^{\widetilde{\pi}(1)} x_{1,d-1} S^{\widetilde{\pi}(d-1)*}    \\ \hline
    \vdots & \vdots & \ddots & \vdots  \\ \hline
    S^{\widetilde{\pi}(d-1)} x_{d-1,0} S^{\widetilde{\pi}(0)*} & S^{\widetilde{\pi}(d-1)} x_{d-1,1} S^{\widetilde{\pi}(1)*} &
    \ldots & S^{\widetilde{\pi}(d-1)} x_{d-1,d-1} S^{\widetilde{\pi}(d-1)*}
         \end{array} \right)\  ,
\end{equation}
\end{widetext}
where
\begin{equation}\label{}
\widetilde{x}_{ij} = \left( \begin{array}{cccc}
    \widetilde{a}^{(0)}_{ij} & \cdot & \ldots & \cdot  \\
    \cdot & \widetilde{a}^{(1)}_{ij} & \ldots & \cdot  \\
    \vdots & \vdots & \ddots & \vdots  \\
    \cdot & \cdot & \ldots & \widetilde{a}^{(d-1)}_{ij} \end{array} \right)\
    .
\end{equation}
 Hence a partial transposition applied to a circulant state
$\rho_\pi$ reduces to
\begin{enumerate}

\item  introducing ``complementary" permutation $\widetilde{\pi}$, and

\item defining a new set of $d \times d$ matrices $\widetilde{a}^{(\alpha)}
=[\widetilde{a}^{(\alpha)}_{ij}]$ .

\end{enumerate}
 Now, ``complementary'' permutation $\widetilde{\pi}$ is fully
characterized by (\ref{pi-pi}). Finally, one finds the following
intricate formula for $\widetilde{a}^{(\alpha)}$:
\begin{equation}\label{a-tilde}
\widetilde{a}^{(\alpha)} \, =\, \sum_{\beta=0}^{d-1}\,
a^{(\alpha+\beta)} \circ
\left(\widetilde{\Pi}\,{S}_\pi^\beta\right)\ , \ \ \ \ \ \ \ \
(\mbox{mod $d$})\ ,
\end{equation}
where ``$\circ$" denotes the Hadamard product, and
\begin{equation}\label{}
    S_\pi = \Pi^* \, S\, \Pi\ .
\end{equation}
 Therefore, we
arrive at our main result

\begin{theorem}
A circulant state $\rho_\pi$ is {\rm PPT} iff the matrices
$\widetilde{a}^{(\alpha)}$ defined in (\ref{a-tilde}) are
positive.
\end{theorem}

\section{Examples} \label{EXAMPLES}

\subsection{PPT class from \cite{III}}

One reconstructs a class of PPT states from \cite{III} taking the
circular decomposition corresponding to trivial permutation with
arbitrary (but positive) $a^{(0)}$ and positive diagonal $a^{(k)}$
$(k=1,\ldots,d-1)$. Note, however, that there are new classes
defined by the same matrices $a^{(\alpha)}$ but corresponding to
different permutations, that is, apart from the state defined by
\begin{equation*}\label{}
    \rho = \sum_{i,j=0}^{d-1} a^{(0)}_{ij} \, e_{ij} \ot e_{ij} +
    \sum_{k=1}^{d-1}\sum_{i=0}^{d-1} a^{(k)}_{ii} \, e_{ii} \ot e_{i+k,i+k} \ ,
\end{equation*}
one has its $\pi$ partner
\begin{eqnarray*}\label{}
    \rho_\pi &=& (\oper \ot \Pi) \rho (\oper \ot \Pi^*) \nonumber
    \\ &=&
\sum_{i,j=0}^{d-1} a^{(0)}_{ij} \, e_{ij} \ot e_{\pi(i),\pi(j)}
\\ &+&
    \sum_{k=1}^{d-1}\sum_{i=0}^{d-1} a^{(k)}_{ii} \, e_{ii} \ot
    e_{\pi(i)+k,\pi(i)+k}\ .
\end{eqnarray*}
It is, therefore, clear that all examples discussed in \cite{III}
(together with the corresponding ``$\pi$--partners'') belong to
our new class.

\subsection{$\pi$--Isotropic state}

The standard isotropic  state \cite{Horodecki} in $d \ot d$
\begin{equation}\label{}
    \mathcal{I}\ =  \frac{1-\lambda}{d^2}\,\mathbb{I} \ot \mathbb{I} + \frac{\lambda}{d}\,
    \sum_{i,j=0}^{d-1}\, e_{ij} \ot e_{ij} \ ,
\end{equation}
corresponds to trivial permutation
 and it is defined by the following set of $d \times d$ positive
 matrices:
 \begin{equation*}\label{}
   a^{(0)}_{ij} = \left\{\begin{array}{ll}
 \lambda/d\ \ & , \ i\neq j \\ \lambda/d + (1-\lambda)/d^2 \ \ &, \ i=j \end{array} \right. \
 ,
\end{equation*}
and diagonal
\begin{equation*}\label{}
   a^{(k)}_{ij} = \left\{\begin{array}{ll}
 0\ \ & , \ i\neq j \\ \lambda/d + (1-\lambda)/d^2 \ \ &, \ i=j \end{array} \right. \
 ,
\end{equation*}
for $k=1,\ldots,d-1\,$. Again, for each permutation $\pi$ we may
define $\pi$--isotropic state
\begin{eqnarray*}\label{}
\mathcal{I} &=& (\oper \ot \Pi) \mathcal{I} (\oper \ot \Pi^*)
\nonumber \\ &=& \frac{1-\lambda}{d^2}\,\mathbb{I} \ot \mathbb{I}
+ \frac{\lambda}{d}\,
    \sum_{i,j=0}^{d-1}\, e_{ij} \ot e_{\pi(i),\pi(j)} \ ,
\end{eqnarray*}
which is defined by the same set of matrices $a^{(\alpha)}$ but
corresponds to $\pi$--decomposition.

\subsection{$\pi$--Werner state}

The celebrated Werner state \cite{Werner} is defined by the
following well known formula
\begin{equation}\label{W}
    \mathcal{W} = (1-p)\, Q^+ + p\,Q^-\ ,
\end{equation}
where
\begin{equation*}\label{}
    Q^\pm = \frac{1}{d(d\pm 1)} \Big( \mathbb{I} \ot \mathbb{I} \pm \mathbb{F} \Big) \
    ,
\end{equation*}
and $\mathbb{F}$ denotes a flip operator defined by
\[   \mathbb{F} = \sum_{i,j=0}^{d-1}\, e_{ij} \ot e_{ji} \ . \]
It is clear that  ${\cal W}$  belongs to a class of bipartite
operators obtained from the class of isotropic states by applying
a partial transposition. One easily finds
\begin{equation*}\label{}
    \widetilde{a}^{(0)}_{ij} = \left\{\begin{array}{ll} x_-\ \ & , \ i\neq j \\
 x_-    + x_+ \ \ &, \ i=j \end{array} \right. \ ,
\end{equation*}
and
\[  \widetilde{a}^{(k)} \ = \ x_+\, \mathbb{I}\ , \ \ \ \ \ k=1,\ldots,d-1\
.\] where
\begin{equation*}\label{}
    x_\pm = \frac{1-p}{d^2 + d} \pm \frac{p}{d^2 - d} \ .
\end{equation*}
It is clear that for any permutation $\pi$ one may define
$\pi$--Werner state
\begin{eqnarray*}\label{}
\mathcal{W}_\pi &=& (\oper \ot \Pi)\, \mathcal{W}\, (\oper \ot
\Pi^*) \nonumber \\ &=& (1-p)\, Q^+_\pi + p\,Q^-_\pi\ , \
    ,
\end{eqnarray*}
where
\begin{equation*}\label{}
    Q^\pm_\pi = \frac{1}{d(d\pm 1)} \Big( \mathbb{I} \ot \mathbb{I} \pm \mathbb{F}_\pi \Big) \
    ,
\end{equation*}
and
 $\mathbb{F}_\pi $ denotes a ``$\pi$--flip operator" defined by
\[   \mathbb{F}_\pi  = (\oper \ot \Pi)\, \mathbb{F}\, (\oper \ot \Pi^*)
=\sum_{i,j=0}^{d-1}\, e_{ij} \ot e_{\pi(j),\pi(j)} \ . \]

\subsection{Ha example in $4 \ot 4$}

 Ha \cite{Ha} constructed a $4\ot 4$ PPT
state which was used to check that the seminal Robertson positive
map $\Lambda : M_4(\mathbb{C}) \longrightarrow M_4(\mathbb{C})$
\cite{Robertson} is indecomposable. Ha's state belongs to our
class labeled by a trivial permutation $\pi_1$ (see Appendix) with
four positive matrices defined as follows:
\[
 a = \left( \begin{array}{cccc}
    1 & \cdot &  -1 & \cdot \\
    \cdot &  \cdot & \cdot & \cdot    \\
    -1 & \cdot & 1 &  \cdot \\
    \cdot & \cdot & \cdot & \cdot  \end{array} \right)
\ , \ \ \ \ \ b=\left( \begin{array}{cccc}
    \cdot & \cdot &  \cdot & \cdot \\
    \cdot &  \cdot & \cdot & \cdot    \\
    \cdot & \cdot & 1 &  \cdot \\
    \cdot & \cdot & \cdot & \cdot  \end{array} \right)\ ,
     \]

\[
 c = \left( \begin{array}{cccc}
    1 & \cdot &  \cdot & \cdot \\
    \cdot &  1 & 1 & \cdot    \\
    \cdot & 1 & 1 &  \cdot \\
    \cdot & \cdot & \cdot & \cdot  \end{array} \right)
\ , \ \ \ \ \ d = \left( \begin{array}{cccc}
    \cdot & \cdot &  \cdot & \cdot \\
    \cdot &  1 & \cdot & \cdot    \\
    \cdot & \cdot & \cdot &  \cdot \\
    \cdot & \cdot & \cdot & \cdot  \end{array} \right) \ ,
     \]
The partially transposed state defines a circulant operator
corresponding to decomposition labeled by $\widetilde{\pi}_1$ and
defined by:
\[
 \widetilde{a}\ =\ \left( \begin{array}{cccc}
    1 & \cdot &  \cdot & \cdot \\
    \cdot &  1 & \cdot & \cdot    \\
    \cdot & \cdot & 1 &  \cdot \\
    \cdot & \cdot & \cdot & \cdot  \end{array} \right)
\ , \ \ \ \ \ \widetilde{b}\ =\ \left( \begin{array}{cccc}
    \cdot & \cdot &  \cdot & \cdot \\
    \cdot &  1 & 1 & \cdot    \\
    \cdot & 1 & 1 &  \cdot \\
    \cdot & \cdot & \cdot & \cdot  \end{array} \right)\ ,
     \]

\[
 \widetilde{c}\ =\ \left( \begin{array}{cccc}
    1 & \cdot & -1 & \cdot \\
    \cdot &  \cdot & \cdot & \cdot    \\
    -1 & \cdot & 1 &  \cdot \\
    \cdot & \cdot & \cdot & \cdot  \end{array} \right)
\ , \ \ \ \ \ \widetilde{d}\ =\ \left( \begin{array}{cccc}
    \cdot & \cdot &  \cdot & \cdot \\
    \cdot &  \cdot & \cdot & \cdot    \\
    \cdot & \cdot & \cdot &  \cdot \\
    \cdot & \cdot & \cdot & \cdot  \end{array} \right) \ .
     \]
Evidently
$\widetilde{a},\widetilde{b},\widetilde{c},\widetilde{d}\geq0 $.
Interestingly, as was shown by Ha \cite{Ha} both $\rho$ and
$\rho^\tau$ are of Schmidt rank two (it proves that Robertson map
is not only indecomposable but even atomic, i.e. it can not be
written as a sum of 2-positive and 2-copositive maps). We stress
that this example does not belong to the previous class defined in
\cite{III}.

\subsection{Fei et. al. bound entangled state in $4 \ot 4$}

Fei et. al. \cite{Fei} constructed $4 \ot 4$ bound entangled state
which correspond to $\widetilde{\pi}_1$--decomposition (see
Appendix). It is defined by the following set of
$\widetilde{a},\widetilde{b},\widetilde{c},\widetilde{d}$:
\[\left( \begin{array}{cccc}
    x_1 & \cdot &  \cdot & \cdot \\
    \cdot &  x_5 & \cdot & -x_5    \\
    \cdot & \cdot & x_1 &  \cdot \\
    \cdot & -x_5 & \cdot & x_5  \end{array} \right)
\ , \ \ \ \ \ \left( \begin{array}{cccc}
    x_3 & -x_3 &  \cdot & \cdot \\
    -x_3 &  x_3 & \cdot & \cdot    \\
    \cdot & \cdot & x_4 &  -x_4 \\
    \cdot & \cdot & -x_4 & x_4  \end{array} \right)\ ,
     \]

\[
 \left( \begin{array}{cccc}
    x_2 & \cdot & -x_2 & \cdot \\
    \cdot &  x_1 & \cdot & \cdot    \\
    -x_2 & \cdot & x_2 &  \cdot \\
    \cdot & \cdot & \cdot & x_1  \end{array} \right)
\ , \ \ \ \ \  \left( \begin{array}{cccc}
    \ \cdot\ & \ \cdot\ &  \ \cdot\ & \ \cdot\ \\
    \, \cdot\, &  \, \cdot\, & \, \cdot\, & \, \cdot\,    \\
    \, \cdot\, & \, \cdot\, & \, \cdot\, &  \, \cdot\, \\
    \, \cdot\, & \, \cdot\, & \, \cdot\, & \, \cdot\,  \end{array} \right) \ .
     \]
Evidently
$\widetilde{a},\widetilde{b},\widetilde{c},\widetilde{d}\geq0 $
for $x_i \geq 0$. Now, partially transposed state is circular with
respect $\pi_1$--decomposition (see Appendix) and it is defined by
the following set of $4 \times 4$ matrices $a,b,c,d$:
\[\left( \begin{array}{cccc}
    x_1 & -x_3 & -x_2 & \cdot \\
    -x_3 &  x_1 & \cdot & -x_5    \\
    -x_2 & \cdot & x_1 &  -x_4 \\
    \cdot & -x_5 & -x_4 & x_1  \end{array} \right)
\ , \ \ \ \ \ \left( \begin{array}{cccc}
    x_3 &\ \cdot\ &\  \cdot\ &\ \cdot\ \\
    \cdot &  \cdot & \cdot & \cdot    \\
    \cdot & \cdot & x_4 &  \cdot \\
    \cdot & \cdot & \cdot & \cdot  \end{array} \right)\ ,
     \]

\[
 \left( \begin{array}{cccc}
      x_2 &\ \cdot\ &\  \cdot\ &\ \cdot\ \\
    \cdot &  x_5 & \cdot & \cdot    \\
    \cdot & \cdot & x_2 &  \cdot \\
    \cdot & \cdot & \cdot & x_5 \end{array} \right)
\ , \ \ \ \ \  \left( \begin{array}{cccc}
    \ \cdot\ &\ \cdot\ &\  \cdot\ &\ \cdot\ \\
    \cdot &  x_3 & \cdot & \cdot    \\
    \cdot & \cdot & \cdot &  \cdot \\
    \cdot & \cdot & \cdot & x_4\end{array} \right) \ .
     \]
It is clear that in general $a$ is not a positive matrix. However,
for $x_1 = (1-\varepsilon)/4$ and $x_2=x_3=x_4=x_5 =
\varepsilon/8$ it has three different eigenvalues
\[ 1/4\ , \ \ (1-2\varepsilon)/4\ , \ \ (1-\varepsilon)/4\ , \]
and hence  $\rho$ is PPT for $0\leq \varepsilon\leq 1/2$. It was
shown \cite{Fei} that $\rho$ being PPT is entangled.

\section{Conclusions}

We have constructed a large class of PPT states in $d \ot d$ which
 correspond to circular decompositions of $\mathbb{C}^d \ot
\mathbb{C}^d$ into direct sums of $d$-dimensional subspaces. This
class significantly enlarges the previous class defined in
\cite{III}. It contains several known examples from the literature
and produces a highly nontrivial family of new states.

There are many open problems: the basic question is how to detect
entanglement within this class of PPT states. One may expect that
there is special class of entanglement witnesses which are
sensitive to entanglement encoded into circular decompositions.
The related mathematical problem is the construction of linear
indecomposable positive maps $\Lambda : M_d(\mathbb{C})
\longrightarrow M_d(\mathbb{C})$ satisfying
\[  (\oper \ot \Lambda)\, \rho \ngeqslant 0\ ,  \]
for some circulant PPT state $\rho$. A corresponding class of such
maps correlated with the previous class of PPT states \cite{III}
was recently proposed in \cite{EW}. It would be interesting to
establish a structure of edge states \cite{edge1,edge2} within
circulant PPT states since  the knowledge of edge states is
sufficient to characterize all PPT states.  Finally, it is
interesting to explore the possibility of other decompositions
leading to new classes of PPT states. We stress that the seminal
Horodecki $3 \ot 3$ entangled PPT state \cite{PPT} does not belong
to our class. In a forthcoming paper we show that this state
belongs to a new class of PPT states which is governed by another
type of decompositions of $\mathbb{C}^d \ot \mathbb{C}^d$.

\section*{Appendix}
\def\theequation{A.\arabic{equation}}
\setcounter{equation}{0}

For $d=4$ one has 6 different  decompositions of $\mathbb{C}^4 \ot
\mathbb{C}^4$ into the direst sum of four 4-dimensional subspaces.
These are labeled by permutations from the symmetric group $S_3$.
One finds
\begin{eqnarray*}\label{}
    (\pi_1(0)=0\, ,\,\pi_1(1)=1\, ,\, \pi_1(2)=2 \, ,\,\pi_1(3)=3) \ ,\\
    (\widetilde{\pi}_1(0)=0 \, ,\,\widetilde{\pi}_1(1)=3 \, ,\,\widetilde{\pi}_1(2)=2\, ,\,\widetilde{\pi}_1(3)=1) \
    ,\\
    (\pi_2(0)=0\, ,\,\pi_2(1)=2\, ,\, \pi_2(2)=3 \, ,\,\pi_3(3)=1) \ ,\\
    (\widetilde{\pi}_2(0)=0 \, ,\,\widetilde{\pi}_2(1)=2 \, ,\,\widetilde{\pi}_2(2)=1\, ,\,\widetilde{\pi}_2(3)=3) \
    ,\\
    (\pi_3(0)=0\, ,\,\pi_3(1)=3\, ,\, \pi_3(2)=1 \, ,\,\pi_3(3)=2) \ ,\\
    (\widetilde{\pi}_3(0)=0 \, ,\,\widetilde{\pi}_3(1)=1 \, ,\,\widetilde{\pi}_3(2)=3\, ,\,\widetilde{\pi}_3(3)=2) \
    .
\end{eqnarray*}
The corresponding permutation matrices read as follows
\begin{equation}\label{}
 \Pi_1 =  \left( \begin{array}{cccc}
    1 & \cdot &  \cdot & \cdot \\
    \cdot &  1 & \cdot & \cdot   \\
    \cdot &  \cdot & 1 & \cdot \\
    \cdot & \cdot & \cdot & 1 \end{array} \right)\  ,\ \ \ \
 \widetilde{\Pi}_1 =  \left( \begin{array}{cccc}
    1 & \cdot &  \cdot & \cdot \\
    \cdot &  \cdot & \cdot & 1  \\
    \cdot &  \cdot & 1 & \cdot \\
    \cdot & 1 & \cdot & \cdot  \end{array} \right)\ ,
\end{equation}
\begin{equation}\label{}
 \Pi_2 = \left( \begin{array}{cccc}
    1 & \cdot &  \cdot & \cdot \\
    \cdot &  \cdot &  \cdot & 1    \\
    \cdot & 1 & \cdot & \cdot  \\
    \cdot & \cdot & 1 & \cdot  \end{array} \right)\  ,\ \ \ \
 \widetilde{\Pi}_2 =  \left( \begin{array}{cccc}
    1 & \cdot &  \cdot & \cdot \\
    \cdot & 1 &  \cdot & \cdot   \\
    \cdot & \cdot  & \cdot & 1 \\
    \cdot &  \cdot & 1 & \cdot  \end{array} \right)\ ,
\end{equation}
\begin{equation}\label{}
 \Pi_3 =   \left( \begin{array}{cccc}
    1 & \cdot &  \cdot & \cdot \\
    \cdot &  \cdot & 1 & \cdot \\
    \cdot & \cdot & \cdot & 1  \\
    \cdot & 1 & \cdot &  \cdot  \end{array} \right)\  ,\ \ \ \
 \widetilde{\Pi}_3 =  \left( \begin{array}{cccc}
    1 & \cdot &  \cdot & \cdot \\
    \cdot &  \cdot & 1 & \cdot   \\
    \cdot & 1 & \cdot  & \cdot  \\
    \cdot &  \cdot & \cdot & 1 \end{array} \right)\ .
\end{equation}
Moreover, one finds for

\begin{itemize}

\item   $\widetilde{\Pi}_1 S_{\pi_1}$, $\widetilde{\Pi}_1 S^2_{\pi_1}$ and
$\widetilde{\Pi}_1 S_{\pi_1}^3$:
\begin{equation*}\label{}
   \left( \begin{array}{cccc}
    \cdot & \cdot &  \cdot & 1 \\
    \cdot &  \cdot & 1 & \cdot   \\
    \cdot &  1 & \cdot & \cdot \\
    1 & \cdot & \cdot & \cdot \end{array} \right)\  ,\ \ \ \
  \left( \begin{array}{cccc}
    \cdot & \cdot &  1 & \cdot \\
    \cdot &  1 & \cdot & \cdot  \\
    1 &  \cdot & \cdot & \cdot \\
    \cdot & \cdot & \cdot & 1  \end{array} \right)\ ,
    \ \ \ \
  \left( \begin{array}{cccc}
    \cdot & 1 &  \cdot & \cdot \\
    1 &  \cdot & \cdot & \cdot  \\
    \cdot &  \cdot & \cdot & 1 \\
    \cdot & \cdot & 1 & \cdot  \end{array} \right)\ ,
\end{equation*}

\item  $\widetilde{\Pi}_2 S_{\pi_2}$, $\widetilde{\Pi}_2 S_{\pi_2}^2$ and
$\widetilde{\Pi}_2 S_{\pi_2}^3$:

\begin{equation*}\label{}
   \left( \begin{array}{cccc}
    \cdot &  \cdot & 1 & \cdot \\
    \cdot &  \cdot & \cdot & 1  \\
    1& \cdot &  \cdot & \cdot  \\
     \cdot & 1 & \cdot & \cdot  \end{array} \right)\  ,\ \ \ \
  \left( \begin{array}{cccc}
    \cdot & 1 & \cdot  & \cdot \\
    1 & \cdot & \cdot & \cdot  \\
     \cdot &  \cdot & 1 & \cdot \\
    \cdot &  \cdot & \cdot & 1  \end{array} \right)\ ,
    \ \ \ \
 \left( \begin{array}{cccc}
    \cdot & \cdot &  \cdot & 1 \\
    \cdot & \cdot & 1 & \cdot   \\
      \cdot & 1 & \cdot &  \cdot \\
    1 & \cdot &\cdot & \cdot \end{array} \right) \ ,
\end{equation*}

\item   and for $\widetilde{\Pi}_3 S_{\pi_3}$, $\widetilde{\Pi}_3 S_{\pi_3}^2$ and
$\widetilde{\Pi}_3 S_{\pi_3}^3$:

\begin{equation*}\label{}
   \left( \begin{array}{cccc}
    \cdot & 1 & \cdot &  \cdot  \\
    \ 1 & \cdot & \cdot & \cdot  \\
    \cdot &  \cdot & \cdot & 1 \\
    \cdot & \cdot & 1 & \cdot \end{array} \right)\  ,\ \ \ \
  \left( \begin{array}{cccc}
    \cdot & \cdot &  \cdot & 1 \\
    \cdot &  1& \cdot &  \cdot   \\
    \cdot & \cdot & 1 & \cdot \\
    1 & \cdot & \cdot & \cdot  \end{array} \right)\ ,
    \ \ \ \
  \left( \begin{array}{cccc}
    \cdot & \cdot & 1 & \cdot \\
    \cdot &  \cdot & \cdot & 1  \\
    1 & \cdot & \cdot & \cdot \\
    \cdot & 1 & \cdot & \cdot \end{array} \right)\ ,
\end{equation*}
respectively.

\end{itemize}

\subsection{$\pi_1$ and $\widetilde{\pi}_1$ circulant states}

\begin{eqnarray*}\label{}
\Sigma^{\pi_1}_0 &=& \mbox{span}\left\{ e_0 \ot e_0\, , e_1 \ot e_1\, ,e_2 \ot e_2\, , e_3 \ot e_3 \right\}
\ ,    \nonumber \\
\Sigma^{\pi_1}_1 &=& \mbox{span}\left\{ e_0 \ot e_1\, , e_1\ot
e_2\, , e_2\ot e_3\, , e_3 \ot e_0 \right\}
 \    ,\\
\Sigma^{\pi_1}_2 &=& \mbox{span}\left\{ e_0 \ot e_2\, , e_1 \ot
e_3\, , e_2\ot e_1\, , e_3 \ot e_2 \right\}
 \    ,\\
\Sigma^{\pi_1}_3 &=& \mbox{span}\left\{ e_0 \ot e_3\, , e_1 \ot
e_0\, , e_2\ot e_1\, , e_3 \ot e_2 \right\} \    ,
\end{eqnarray*}

\begin{eqnarray*}\label{}
\widetilde{\Sigma}^{\pi_1}_0 &=& \mbox{span}\left\{ e_0 \ot e_0\,
, e_1 \ot e_3\, ,e_2 \ot e_2\, , e_3 \ot e_1 \right\}
\ ,    \nonumber \\
\widetilde{\Sigma}^{\pi_1}_1 &=& \mbox{span}\left\{ e_0 \ot e_1\,
, e_1 \ot e_0\, ,e_2 \ot e_3\, , e_3 \ot e_2 \right\}
 \    ,\\
\widetilde{\Sigma}^{\pi_1}_2 &=& \mbox{span}\left\{ e_0 \ot e_2\,
, e_1 \ot e_1\, ,e_2 \ot e_0\, , e_3 \ot e_3 \right\}
 \    ,\\
\widetilde{\Sigma}^{\pi_1}_3 &=& \mbox{span}\left\{ e_0 \ot e_3\,
, e_1 \ot e_2\, ,e_2 \ot e_1\, , e_3 \ot e_0 \right\} \    .
\end{eqnarray*}

\begin{widetext}
\begin{equation}\label{4C1}
  \rho_{\pi_1}\ =\ \left( \begin{array}{cccc|cccc|cccc|cccc}
    a_{00} & \cdot  & \cdot  & \cdot  &
    \cdot  & a_{01} & \cdot  & \cdot  &
    \cdot  & \cdot  & a_{02} & \cdot  &
    \cdot  & \cdot  & \cdot  & a_{03}   \\
    \cdot  & b_{00} & \cdot  & \cdot  &
    \cdot  & \cdot  & b_{01} & \cdot  &
    \cdot  & \cdot  & \cdot  & b_{02} &
    b_{03} & \cdot  & \cdot  & \cdot    \\
    \cdot  & \cdot  & c_{00} & \cdot  &
    \cdot  & \cdot  & \cdot  & c_{01} &
    c_{02} & \cdot  & \cdot  & \cdot  &
    \cdot  & c_{03} & \cdot  & \cdot     \\
    \cdot  & \cdot  & \cdot  & d_{00} &
    d_{01} & \cdot  & \cdot  & \cdot  &
    \cdot  & d_{02} & \cdot  & \cdot  &
    \cdot  & \cdot  & d_{03} & \cdot     \\   \hline
    %%%%%%%%%%%%%%%%%%%%%%%%%%%%%%%%%%%%%%%%%%
    \cdot  & \cdot  & \cdot  & d_{10} &
    d_{11} & \cdot  & \cdot  & \cdot  &
    \cdot  & d_{12} & \cdot  & \cdot  &
    \cdot  & \cdot  & d_{13} & \cdot     \\
    a_{10} & \cdot  & \cdot  & \cdot  &
    \cdot  & a_{11} & \cdot  & \cdot  &
    \cdot  & \cdot  & a_{12} & \cdot  &
    \cdot  & \cdot  & \cdot  & a_{13}   \\
    \cdot  & b_{10} & \cdot  & \cdot  &
    \cdot  & \cdot  & b_{11} & \cdot  &
    \cdot  & \cdot  & \cdot  & b_{12} &
    b_{13} & \cdot  & \cdot  & \cdot    \\
    \cdot  & \cdot  & c_{10} & \cdot  &
    \cdot  & \cdot  & \cdot  & c_{11} &
    c_{12} & \cdot  & \cdot  & \cdot  &
    \cdot  & c_{13} & \cdot  & \cdot    \\   \hline
    %%%%%%%%%%%%%%%%%%%%%%%%%%%%%%%%%%%%%%%%%%
    \cdot  & \cdot  & c_{20} & \cdot  &
    \cdot  & \cdot  & \cdot  & c_{21} &
    c_{22} & \cdot  & \cdot  & \cdot  &
    \cdot  & c_{23} & \cdot  & \cdot    \\
    \cdot  & \cdot  & \cdot  & d_{20} &
    d_{21} & \cdot  & \cdot  & \cdot  &
    \cdot  & d_{22} & \cdot  & \cdot  &
    \cdot  & \cdot  & d_{23} & \cdot     \\
    a_{20} & \cdot  & \cdot  & \cdot  &
    \cdot  & a_{21} & \cdot  & \cdot  &
    \cdot  & \cdot  & a_{22} & \cdot  &
    \cdot  & \cdot  & \cdot  & a_{23}   \\
    \cdot  & b_{20} & \cdot  & \cdot  &
    \cdot  & \cdot  & b_{21} & \cdot  &
    \cdot  & \cdot  & \cdot  & b_{22} &
    b_{23} & \cdot  & \cdot  & \cdot    \\   \hline
    %%%%%%%%%%%%%%%%%%%%%%%%%%%%%%%%%%%%%%%%%%
    \cdot  & b_{30} & \cdot  & \cdot  &
    \cdot  & \cdot  & b_{31} & \cdot  &
    \cdot  & \cdot  & \cdot  & b_{32} &
    b_{33} & \cdot  & \cdot  & \cdot    \\
    \cdot  & \cdot  & c_{30} & \cdot  &
    \cdot  & \cdot  & \cdot  & c_{31} &
    c_{32} & \cdot  & \cdot  & \cdot  &
    \cdot  & c_{33} & \cdot  & \cdot    \\
    \cdot  & \cdot  & \cdot  & d_{30} &
    d_{31} & \cdot  & \cdot  & \cdot  &
    \cdot  & d_{32} & \cdot  & \cdot  &
    \cdot  & \cdot  & d_{33} & \cdot     \\
    a_{30} & \cdot  & \cdot  & \cdot  &
    \cdot  & a_{31} & \cdot  & \cdot  &
    \cdot  & \cdot  & a_{32} & \cdot  &
    \cdot  & \cdot  & \cdot  & a_{33}
     \end{array} \right)\ .
\end{equation}

\begin{equation}\label{4C-T}
  \rho^\tau_{\pi_1}\ =\ \left( \begin{array}{cccc|cccc|cccc|cccc}
    \widetilde{a}_{00} & \cdot  & \cdot  & \cdot  &
    \cdot  & \cdot  & \cdot  & \widetilde{a}_{01} &
    \cdot  & \cdot  & \widetilde{a}_{02} & \cdot  &
    \cdot  & \widetilde{a}_{03} & \cdot  & \cdot    \\
    \cdot  & \widetilde{b}_{00} & \cdot  & \cdot  &
    \widetilde{b}_{01} & \cdot  & \cdot  & \cdot  &
    \cdot  & \cdot  & \cdot  & \widetilde{b}_{02} &
    \cdot  & \cdot  & \widetilde{b}_{03} & \cdot    \\
    \cdot  & \cdot  & \widetilde{c}_{00} & \cdot  &
    \cdot  & \widetilde{c}_{01} & \cdot  & \cdot  &
    \widetilde{c}_{02} & \cdot  & \cdot  & \cdot  &
    \cdot  & \cdot  & \cdot  & \widetilde{c}_{03}    \\
    \cdot  & \cdot  & \cdot  & \widetilde{d}_{00} &
    \cdot  & \cdot  & \widetilde{d}_{01} & \cdot  &
    \cdot  & \widetilde{d}_{02} & \cdot  & \cdot  &
    \widetilde{d}_{03} & \cdot  & \cdot  & \cdot     \\   \hline
    %%%%%%%%%%%%%%%%%%%%%%%%%%%%%%%%%%%%%%%%%%
    \cdot  & \widetilde{b}_{10} & \cdot  & \cdot  &
    \widetilde{b}_{11} & \cdot  & \cdot  & \cdot  &
    \cdot  & \cdot  & \cdot  & \widetilde{b}_{12} &
    \cdot  & \cdot  & \widetilde{b}_{13} & \cdot    \\
    \cdot  & \cdot  & \widetilde{c}_{10} & \cdot  &
    \cdot  & \widetilde{c}_{11} & \cdot  & \cdot  &
    \widetilde{c}_{12} & \cdot  & \cdot  & \cdot  &
    \cdot  & \cdot  & \cdot  & \widetilde{c}_{13}    \\
    \cdot  & \cdot  & \cdot  & \widetilde{d}_{10} &
    \cdot  & \cdot  & \widetilde{d}_{11} & \cdot  &
    \cdot  & \widetilde{d}_{12} & \cdot  & \cdot  &
    \widetilde{d}_{13} & \cdot  & \cdot  & \cdot     \\
    \widetilde{a}_{10} & \cdot  & \cdot  & \cdot  &
    \cdot  & \cdot  & \cdot  & \widetilde{a}_{11} &
    \cdot  & \cdot  & \widetilde{a}_{12} & \cdot  &
    \cdot  & \widetilde{a}_{13} & \cdot  & \cdot     \\   \hline
    %%%%%%%%%%%%%%%%%%%%%%%%%%%%%%%%%%%%%%%%%%
    \cdot  & \cdot  & \widetilde{c}_{20} & \cdot  &
    \cdot  & \widetilde{c}_{21} & \cdot  & \cdot  &
    \widetilde{c}_{22} & \cdot  & \cdot  & \cdot  &
    \cdot  & \cdot  & \cdot  & \widetilde{c}_{23}    \\
    \cdot  & \cdot  & \cdot  & \widetilde{d}_{20} &
    \cdot  & \cdot  & \widetilde{d}_{21} & \cdot  &
    \cdot  & \widetilde{d}_{22} & \cdot  & \cdot  &
    \widetilde{d}_{23} & \cdot  & \cdot  & \cdot     \\
    \widetilde{a}_{20} & \cdot  & \cdot  & \cdot  &
    \cdot  & \cdot  & \cdot  & \widetilde{a}_{21} &
    \cdot  & \cdot  & \widetilde{a}_{22} & \cdot  &
    \cdot  & \widetilde{a}_{23} & \cdot  & \cdot    \\
    \cdot  & \widetilde{b}_{20} & \cdot  & \cdot  &
    \widetilde{b}_{21} & \cdot  & \cdot  & \cdot  &
    \cdot  & \cdot  & \cdot  & \widetilde{b}_{22} &
    \cdot  & \cdot  & \widetilde{b}_{23} & \cdot    \\   \hline
    %%%%%%%%%%%%%%%%%%%%%%%%%%%%%%%%%%%%%%%%%%
    \cdot  & \cdot  & \cdot  & \widetilde{d}_{30} &
    \cdot  & \cdot  & \widetilde{d}_{31} & \cdot  &
    \cdot  & \widetilde{d}_{32} & \cdot  & \cdot  &
    \widetilde{d}_{33} & \cdot  & \cdot  & \cdot     \\
    \widetilde{a}_{30} & \cdot  & \cdot  & \cdot  &
    \cdot  & \cdot  & \cdot  & \widetilde{a}_{31} &
    \cdot  & \cdot  & \widetilde{a}_{32} & \cdot  &
    \cdot  & \widetilde{a}_{33} & \cdot  & \cdot    \\
    \cdot  & \widetilde{b}_{30} & \cdot  & \cdot  &
    \widetilde{b}_{31} & \cdot  & \cdot  & \cdot  &
    \cdot  & \cdot  & \cdot  & \widetilde{b}_{32} &
    \cdot  & \cdot  & \widetilde{b}_{33} & \cdot    \\
    \cdot  & \cdot  & \widetilde{c}_{30} & \cdot  &
    \cdot  & \widetilde{c}_{31} & \cdot  & \cdot  &
    \widetilde{c}_{32} & \cdot  & \cdot  & \cdot  &
    \cdot  & \cdot  & \cdot  & \widetilde{c}_{33}
\end{array} \right)\ .
\end{equation}
\end{widetext}
where the matrices
$\widetilde{a},\widetilde{b},\widetilde{c},\widetilde{d}$ are
given by
\begin{widetext}
%\begin{equation*}\label{}
\[ \widetilde{a}\ = \ \left( \begin{array}{cccc}
    a_{00} & d_{01} & c_{02} & b_{03} \\
    d_{10} & c_{11} & b_{12} & a_{13}   \\
    c_{20} & b_{21} & a_{22} & d_{23} \\
    b_{30} & a_{31} & d_{32} & c_{33} \end{array} \right)\ , \ \
\widetilde{b}\ = \ \left( \begin{array}{cccc}
    b_{00} & a_{01} & d_{02} & c_{03} \\
    a_{10} & d_{11} & c_{12} & b_{13}   \\
    d_{20} & c_{21} & b_{22} & a_{23} \\
    c_{30} & b_{31} & a_{32} & d_{33}\end{array} \right)\ , \ \
%\end{equation*}
%\begin{equation*}\label{}
\widetilde{c}\ = \ \left( \begin{array}{cccc}
    c_{00} & b_{01} & a_{02} & d_{03} \\
    b_{10} & a_{11} & d_{12} & c_{13}   \\
    a_{20} & d_{21} & c_{22} & b_{23} \\
    d_{30} & c_{31} & b_{32} & a_{33}\end{array} \right)\ , \ \
\widetilde{d}\ = \ \left( \begin{array}{cccc}
    d_{00} & c_{01} & b_{02} & a_{03} \\
    c_{10} & b_{11} & a_{12} & d_{13}   \\
    b_{20} & a_{21} & d_{22} & c_{23} \\
    a_{30} & d_{31} & c_{32} & b_{33}\end{array} \right)\ .
%\end{equation*}
\]
\end{widetext}

\subsection{$\pi_2$ and $\widetilde{\pi}_2$ circulant states}

\begin{eqnarray*}\label{}
\Sigma^{\pi_2}_0 &=& \mbox{span}\left\{ e_0 \ot e_0\, , e_1 \ot
e_2\, ,e_2 \ot e_3\, , e_3 \ot e_1 \right\}
\ ,    \nonumber \\
\Sigma^{\pi_2}_1 &=& \mbox{span}\left\{ e_0 \ot e_1\, , e_1 \ot
e_3\, ,e_2 \ot e_0\, , e_3 \ot e_2 \right\}
 \    ,\\
\Sigma^{\pi_2}_2 &=& \mbox{span}\left\{ e_0 \ot e_2\, , e_1 \ot
e_0\, ,e_2 \ot e_1\, , e_3 \ot e_3 \right\}
 \    ,\\
\Sigma^{\pi_2}_3 &=&  \mbox{span}\left\{ e_0 \ot e_3\, , e_1 \ot
e_1\, ,e_2 \ot e_2\, , e_3 \ot e_0 \right\} \    ,
\end{eqnarray*}

\begin{eqnarray*}\label{}
\widetilde{\Sigma}^{\pi_2}_0 &=& \mbox{span}\left\{ e_0 \ot e_0\,
, e_1 \ot e_2\, ,e_2 \ot e_1\, , e_3 \ot e_3 \right\}
\ ,    \nonumber \\
\widetilde{\Sigma}^{\pi_2}_1 &=& \mbox{span}\left\{ e_0 \ot e_1\,
, e_1 \ot e_3\, ,e_2 \ot e_2\, , e_3 \ot e_0 \right\}
 \    ,\\
\widetilde{\Sigma}^{\pi_2}_2 &=& \mbox{span}\left\{ e_0 \ot e_2\,
, e_1 \ot e_0\, ,e_2 \ot e_3\, , e_3 \ot e_1 \right\}
 \    ,\\
\widetilde{\Sigma}^{\pi_2}_3 &=&  \mbox{span}\left\{ e_0 \ot e_3\,
, e_1 \ot e_1\, ,e_2 \ot e_0\, , e_3 \ot e_2 \right\} \    .
\end{eqnarray*}

\begin{widetext}
\begin{equation}\label{4C2}
  \rho_{\pi_2}\ =\ \left( \begin{array}{cccc|cccc|cccc|cccc}
    a_{00} & \cdot  & \cdot  & \cdot  &
    \cdot  & \cdot  & a_{01} & \cdot  &
    \cdot  & \cdot  & \cdot  & a_{02} &
    \cdot  & a_{03} & \cdot  & \cdot      \\
    \cdot  & b_{00} & \cdot  & \cdot  &
    \cdot  & \cdot  & \cdot  & b_{01} &
    b_{02} & \cdot  & \cdot  & \cdot  &
    \cdot  & \cdot  & b_{03} & \cdot          \\
    \cdot  & \cdot  & c_{00} & \cdot  &
    c_{01} & \cdot  & \cdot  & \cdot  &
    \cdot  & c_{02} & \cdot  & \cdot  &
    \cdot  & \cdot  & \cdot  & c_{03}         \\
    \cdot  & \cdot  & \cdot  & d_{00} &
    \cdot  & d_{01} & \cdot  & \cdot  &
    \cdot  & \cdot  & d_{02} & \cdot  &
    d_{03} & \cdot  & \cdot  & \cdot          \\   \hline
    %%%%%%%%%%%%%%%%%%%%%%%%%%%%%%%%%%%%%%%%%%
    \cdot  & \cdot  & c_{10} & \cdot  &
    c_{11} & \cdot  & \cdot  & \cdot  &
    \cdot  & c_{12} & \cdot  & \cdot  &
    \cdot  & \cdot  & \cdot  & c_{13}         \\
    \cdot  & \cdot  & \cdot  & d_{10} &
    \cdot  & d_{11} & \cdot  & \cdot  &
    \cdot  & \cdot  & d_{12} & \cdot  &
    d_{13} & \cdot  & \cdot  & \cdot          \\
    a_{10} & \cdot  & \cdot  & \cdot  &
    \cdot  & \cdot  & a_{11} & \cdot  &
    \cdot  & \cdot  & \cdot  & a_{12} &
    \cdot  & a_{13} & \cdot  & \cdot          \\
    \cdot  & b_{10} & \cdot  & \cdot  &
    \cdot  & \cdot  & \cdot  & b_{11} &
    b_{12} & \cdot  & \cdot  & \cdot  &
    \cdot  & \cdot  & b_{13} & \cdot          \\   \hline
    %%%%%%%%%%%%%%%%%%%%%%%%%%%%%%%%%%%%%%%%%%
    \cdot  & b_{20} & \cdot  & \cdot  &
    \cdot  & \cdot  & \cdot  & b_{21} &
    b_{22} & \cdot  & \cdot  & \cdot  &
    \cdot  & \cdot  & b_{23} & \cdot          \\
    \cdot  & \cdot  & c_{20} & \cdot  &
    c_{21} & \cdot  & \cdot  & \cdot  &
    \cdot  & c_{22} & \cdot  & \cdot  &
    \cdot  & \cdot  & \cdot  & c_{23}         \\
    \cdot  & \cdot  & \cdot  & d_{20} &
    \cdot  & d_{21} & \cdot  & \cdot  &
    \cdot  & \cdot  & d_{22} & \cdot  &
    d_{23} & \cdot  & \cdot  & \cdot          \\
    a_{20} & \cdot  & \cdot  & \cdot  &
    \cdot  & \cdot  & a_{21} & \cdot  &
    \cdot  & \cdot  & \cdot  & a_{22} &
    \cdot  & a_{23} & \cdot  & \cdot        \\   \hline
    %%%%%%%%%%%%%%%%%%%%%%%%%%%%%%%%%%%%%%%%%%
    \cdot  & \cdot  & \cdot  & d_{30} &
    \cdot  & d_{31} & \cdot  & \cdot  &
    \cdot  & \cdot  & d_{32} & \cdot  &
    d_{33} & \cdot  & \cdot  & \cdot          \\
    a_{30} & \cdot  & \cdot  & \cdot  &
    \cdot  & \cdot  & a_{31} & \cdot  &
    \cdot  & \cdot  & \cdot  & a_{32} &
    \cdot  & a_{33} & \cdot  & \cdot          \\
    \cdot  & b_{30} & \cdot  & \cdot  &
    \cdot  & \cdot  & \cdot  & b_{31} &
    b_{32} & \cdot  & \cdot  & \cdot  &
    \cdot  & \cdot  & b_{33} & \cdot         \\
    \cdot  & \cdot  & c_{30} & \cdot  &
    c_{31} & \cdot  & \cdot  & \cdot  &
    \cdot  & c_{32} & \cdot  & \cdot  &
    \cdot  & \cdot  & \cdot  & c_{33}
\end{array} \right)\ .
\end{equation}

\begin{equation}\label{4C2-T}
  \rho^\tau_{\pi_2}\ =\ \left( \begin{array}{cccc|cccc|cccc|cccc}
    \widetilde{a}_{00} & \cdot  & \cdot  & \cdot  &
    \cdot  & \cdot  & \widetilde{a}_{01} & \cdot  &
    \cdot  & \widetilde{a}_{02} & \cdot  & \cdot  &
    \cdot  & \cdot  & \cdot  & \widetilde{a}_{03}     \\
    \cdot  & \widetilde{b}_{00} & \cdot  & \cdot  &
    \cdot  & \cdot  & \cdot  & \widetilde{b}_{01} &
    \cdot  & \cdot  & \widetilde{b}_{02} & \cdot  &
    \widetilde{b}_{03} & \cdot  & \cdot  & \cdot      \\
    \cdot  & \cdot  & \widetilde{c}_{00} & \cdot  &
    \widetilde{c}_{01} & \cdot  & \cdot  & \cdot  &
    \cdot  & \cdot  & \cdot  & \widetilde{c}_{02} &
    \cdot  & \widetilde{c}_{03} & \cdot  & \cdot      \\
    \cdot  & \cdot  & \cdot  & \widetilde{d}_{00} &
    \cdot  & \widetilde{d}_{01} & \cdot  & \cdot  &
    \widetilde{d}_{02} & \cdot  & \cdot  & \cdot  &
    \cdot  & \cdot  & \widetilde{d}_{03} & \cdot      \\   \hline
    %%%%%%%%%%%%%%%%%%%%%%%%%%%%%%%%%%%%%%%%%%
    \cdot  & \cdot  & \widetilde{c}_{10} & \cdot  &
    \widetilde{c}_{11} & \cdot  & \cdot  & \cdot  &
    \cdot  & \cdot  & \cdot  & \widetilde{c}_{12} &
    \cdot  & \widetilde{c}_{13} & \cdot  & \cdot      \\
    \cdot  & \cdot  & \cdot  & \widetilde{d}_{10} &
    \cdot  & \widetilde{d}_{11} & \cdot  & \cdot  &
    \widetilde{d}_{12} & \cdot  & \cdot  & \cdot  &
    \cdot  & \cdot  & \widetilde{d}_{13} & \cdot      \\
    \widetilde{a}_{10} & \cdot  & \cdot  & \cdot  &
    \cdot  & \cdot  & \widetilde{a}_{11} & \cdot  &
    \cdot  & \widetilde{a}_{12} & \cdot  & \cdot  &
    \cdot  & \cdot  & \cdot  & \widetilde{a}_{13}     \\
    \cdot  & \widetilde{b}_{10} & \cdot  & \cdot  &
    \cdot  & \cdot  & \cdot  & \widetilde{b}_{11} &
    \cdot  & \cdot  & \widetilde{b}_{12} & \cdot  &
    \widetilde{b}_{13} & \cdot  & \cdot  & \cdot      \\   \hline
    %%%%%%%%%%%%%%%%%%%%%%%%%%%%%%%%%%%%%%%%%%
    \cdot  & \cdot  & \cdot  & \widetilde{d}_{20} &
    \cdot  & \widetilde{d}_{21} & \cdot  & \cdot  &
    \widetilde{d}_{22} & \cdot  & \cdot  & \cdot  &
    \cdot  & \cdot  & \widetilde{d}_{23} & \cdot      \\
    \widetilde{a}_{20} & \cdot  & \cdot  & \cdot  &
    \cdot  & \cdot  & \widetilde{a}_{21} & \cdot  &
    \cdot  & \widetilde{a}_{22} & \cdot  & \cdot  &
    \cdot  & \cdot  & \cdot  & \widetilde{a}_{23}     \\
    \cdot  & \widetilde{b}_{20} & \cdot  & \cdot  &
    \cdot  & \cdot  & \cdot  & \widetilde{b}_{21} &
    \cdot  & \cdot  & \widetilde{b}_{22} & \cdot  &
    \widetilde{b}_{23} & \cdot  & \cdot  & \cdot      \\
    \cdot  & \cdot  & \widetilde{c}_{20} & \cdot  &
    \widetilde{c}_{21} & \cdot  & \cdot  & \cdot  &
    \cdot  & \cdot  & \cdot  & \widetilde{c}_{22} &
    \cdot  & \widetilde{c}_{23} & \cdot  & \cdot      \\   \hline
    %%%%%%%%%%%%%%%%%%%%%%%%%%%%%%%%%%%%%%%%%%
    \cdot  & \widetilde{b}_{30} & \cdot  & \cdot  &
    \cdot  & \cdot  & \cdot  & \widetilde{b}_{31} &
    \cdot  & \cdot  & \widetilde{b}_{32} & \cdot  &
    \widetilde{b}_{33} & \cdot  & \cdot  & \cdot      \\
    \cdot  & \cdot  & \widetilde{c}_{30} & \cdot  &
    \widetilde{c}_{31} & \cdot  & \cdot  & \cdot  &
    \cdot  & \cdot  & \cdot  & \widetilde{c}_{32} &
    \cdot  & \widetilde{c}_{33} & \cdot  & \cdot      \\
    \cdot  & \cdot  & \cdot  & \widetilde{d}_{30} &
    \cdot  & \widetilde{d}_{31} & \cdot  & \cdot  &
    \widetilde{d}_{32} & \cdot  & \cdot  & \cdot  &
    \cdot  & \cdot  & \widetilde{d}_{33} & \cdot      \\
    \widetilde{a}_{30} & \cdot  & \cdot  & \cdot  &
    \cdot  & \cdot  & \widetilde{a}_{31} & \cdot  &
    \cdot  & \widetilde{a}_{32} & \cdot  & \cdot  &
    \cdot  & \cdot  & \cdot  & \widetilde{a}_{33}
\end{array} \right)\ .
\end{equation}

\end{widetext}
where the matrices
$\widetilde{a},\widetilde{b},\widetilde{c},\widetilde{d}$ are
given by
\begin{widetext}
\[
%\begin{equation*}\label{}
\widetilde{a}\ = \ \left( \begin{array}{cccc}
    a_{00} & c_{01} & b_{02} & d_{03} \\
    c_{10} & a_{11} & d_{12} & b_{13}   \\
    b_{20} & d_{21} & c_{22} & a_{23} \\
    d_{30} & b_{31} & a_{32} & c_{33} \end{array} \right)\ , \ \
    \widetilde{b}\ = \ \left( \begin{array}{cccc}
    b_{00} & d_{01} & c_{02} & a_{03} \\
    d_{10} & b_{11} & a_{12} & c_{13}   \\
    c_{20} & a_{21} & d_{22} & b_{23} \\
    a_{30} & c_{31} & b_{32} & d_{33}\end{array} \right)\ , \ \
%\end{equation*}
%\begin{equation*}\label{}
    \widetilde{c}\ = \ \left( \begin{array}{cccc}
    c_{00} & a_{01} & d_{02} & b_{03} \\
    a_{10} & c_{11} & b_{12} & d_{13}   \\
    d_{20} & b_{21} & a_{22} & c_{23} \\
    b_{30} & d_{31} & c_{32} & a_{33}\end{array} \right)\ , \ \
    \widetilde{d}\ = \ \left( \begin{array}{cccc}
    d_{00} & b_{01} & a_{02} & c_{03} \\
    b_{10} & d_{11} & c_{12} & a_{13}   \\
    a_{20} & c_{21} & b_{22} & d_{23} \\
    c_{30} & a_{31} & d_{32} & b_{33} \end{array} \right)\ .
%\end{equation*}
\]
\end{widetext}

%\newpage

\subsection{$\pi_3$ and $\widetilde{\pi}_3$ circulant states}

\begin{eqnarray*}\label{}
\Sigma^{\pi_3}_0 &=& \mbox{span}\left\{ e_0 \ot e_0\, , e_1 \ot
e_3\, ,e_2 \ot e_1\, , e_3 \ot e_2 \right\}
\ ,    \nonumber \\
\Sigma^{\pi_3}_1 &=& \mbox{span}\left\{ e_0 \ot e_1\, , e_1 \ot
e_0\, ,e_2 \ot e_2\, , e_3 \ot e_3 \right\}
 \    ,\\
\Sigma^{\pi_3}_2 &=& \mbox{span}\left\{ e_0 \ot e_2\, , e_1 \ot
e_1\, ,e_2 \ot e_3\, , e_3 \ot e_0 \right\}
 \    ,\\
\Sigma^{\pi_3}_3 &=&  \mbox{span}\left\{ e_0 \ot e_3\, , e_1 \ot
e_2\, ,e_2 \ot e_0\, , e_3 \ot e_1 \right\} \    ,
\end{eqnarray*}

\begin{eqnarray*}\label{}
\widetilde{\Sigma}^{\pi_3}_0 &=& \mbox{span}\left\{ e_0 \ot e_0\,
, e_1 \ot e_1\, ,e_2 \ot e_3\, , e_3 \ot e_2 \right\}
\ ,    \nonumber \\
\widetilde{\Sigma}^{\pi_3}_1 &=& \mbox{span}\left\{ e_0 \ot e_1\,
, e_1 \ot e_2\, ,e_2 \ot e_0\, , e_3 \ot e_3 \right\}
 \    ,\\
\widetilde{\Sigma}^{\pi_3}_2 &=& \mbox{span}\left\{ e_0 \ot e_2\,
, e_1 \ot e_3\, ,e_2 \ot e_1\, , e_3 \ot e_0 \right\}
 \    ,\\
\widetilde{\Sigma}^{\pi_3}_3 &=&  \mbox{span}\left\{ e_0 \ot e_3\,
, e_1 \ot e_0\, ,e_2 \ot e_2\, , e_3 \ot e_1 \right\} \    .
\end{eqnarray*}

\begin{widetext}
\begin{equation}\label{4C3}
  \rho_{\pi_3}\ =\ \left( \begin{array}{cccc|cccc|cccc|cccc}
    a_{00} & \cdot  & \cdot  & \cdot  &
    \cdot  & \cdot  & \cdot  & a_{01} &
    \cdot  & a_{02} & \cdot  & \cdot  &
    \cdot  & \cdot  & a_{03} & \cdot      \\
    \cdot  & b_{00} & \cdot  & \cdot  &
    b_{01} & \cdot  & \cdot  & \cdot  &
    \cdot  & \cdot  & b_{02} & \cdot  &
    \cdot  & \cdot  & \cdot  & b_{03}    \\
    \cdot  & \cdot  & c_{00} & \cdot  &
    \cdot  & c_{01} & \cdot  & \cdot  &
    \cdot  & \cdot  & \cdot  & c_{02} &
    c_{03} & \cdot  & \cdot  & \cdot     \\
    \cdot  & \cdot  & \cdot  & d_{00} &
    \cdot  & \cdot  & d_{01} & \cdot  &
    d_{02} & \cdot  & \cdot  & \cdot  &
    \cdot  & d_{03} & \cdot  & \cdot      \\   \hline
    %%%%%%%%%%%%%%%%%%%%%%%%%%%%%%%%%%%%%%%%%%
    \cdot  & b_{10} & \cdot  & \cdot  &
    b_{11} & \cdot  & \cdot  & \cdot  &
    \cdot  & \cdot  & b_{12} & \cdot  &
    \cdot  & \cdot  & \cdot  & b_{13}    \\
    \cdot  & \cdot  & c_{10} & \cdot  &
    \cdot  & c_{11} & \cdot  & \cdot  &
    \cdot  & \cdot  & \cdot  & c_{12} &
    c_{13} & \cdot  & \cdot  & \cdot     \\
    \cdot  & \cdot  & \cdot  & d_{10} &
    \cdot  & \cdot  & d_{11} & \cdot  &
    d_{12} & \cdot  & \cdot  & \cdot  &
    \cdot  & d_{13} & \cdot  & \cdot     \\
    a_{10} & \cdot  & \cdot  & \cdot  &
    \cdot  & \cdot  & \cdot  & a_{11} &
    \cdot  & a_{12} & \cdot  & \cdot  &
    \cdot  & \cdot  & a_{13} & \cdot     \\   \hline
    %%%%%%%%%%%%%%%%%%%%%%%%%%%%%%%%%%%%%%%%%%
    \cdot  & \cdot  & \cdot  & d_{20} &
    \cdot  & \cdot  & d_{21} & \cdot  &
    d_{22} & \cdot  & \cdot  & \cdot  &
    \cdot  & d_{23} & \cdot  & \cdot     \\
    a_{20} & \cdot  & \cdot  & \cdot  &
    \cdot  & \cdot  & \cdot  & a_{21} &
    \cdot  & a_{22} & \cdot  & \cdot  &
    \cdot  & \cdot  & a_{23} & \cdot      \\
    \cdot  & b_{20} & \cdot  & \cdot  &
    b_{21} & \cdot  & \cdot  & \cdot  &
    \cdot  & \cdot  & b_{22} & \cdot  &
    \cdot  & \cdot  & \cdot  & b_{23}    \\
    \cdot  & \cdot  & c_{20} & \cdot  &
    \cdot  & c_{21} & \cdot  & \cdot  &
    \cdot  & \cdot  & \cdot  & c_{22} &
    c_{23} & \cdot  & \cdot  & \cdot      \\   \hline
    %%%%%%%%%%%%%%%%%%%%%%%%%%%%%%%%%%%%%%%%%%
    \cdot  & \cdot  & c_{30} & \cdot  &
    \cdot  & c_{31} & \cdot  & \cdot  &
    \cdot  & \cdot  & \cdot  & c_{32} &
    c_{33} & \cdot  & \cdot  & \cdot     \\
    \cdot  & \cdot  & \cdot  & d_{30} &
    \cdot  & \cdot  & d_{31} & \cdot  &
    d_{32} & \cdot  & \cdot  & \cdot  &
    \cdot  & d_{33} & \cdot  & \cdot     \\
    a_{30} & \cdot  & \cdot  & \cdot  &
    \cdot  & \cdot  & \cdot  & a_{31} &
    \cdot  & a_{32} & \cdot  & \cdot  &
    \cdot  & \cdot  & a_{33} & \cdot      \\
    \cdot  & b_{30} & \cdot  & \cdot  &
    b_{31} & \cdot  & \cdot  & \cdot  &
    \cdot  & \cdot  & b_{32} & \cdot  &
    \cdot  & \cdot  & \cdot  & b_{33}
\end{array} \right)\ .
\end{equation}

\begin{equation}\label{4C3-T}
  \rho^\tau_{\pi_3}\ =\ \left( \begin{array}{cccc|cccc|cccc|cccc}
    \widetilde{a}_{00} & \cdot  & \cdot  & \cdot  &
    \cdot  & \widetilde{a}_{01} & \cdot  & \cdot  &
    \cdot  & \cdot  & \cdot  & \widetilde{a}_{02} &
    \cdot  & \cdot  & \widetilde{a}_{03} & \cdot      \\
    \cdot  & \widetilde{b}_{00} & \cdot  & \cdot  &
    \cdot  & \cdot  & \widetilde{b}_{01} & \cdot  &
    \widetilde{b}_{02} & \cdot  & \cdot  & \cdot  &
    \cdot  & \cdot  & \cdot  & \widetilde{b}_{03}    \\
    \cdot  & \cdot  & \widetilde{c}_{00} & \cdot  &
    \cdot  & \cdot  & \cdot  & \widetilde{c}_{01} &
    \cdot  & \widetilde{c}_{02} & \cdot  & \cdot  &
    \widetilde{c}_{03} & \cdot  & \cdot  & \cdot     \\
    \cdot  & \cdot  & \cdot  & \widetilde{d}_{00} &
    \widetilde{d}_{01} & \cdot  & \cdot  & \cdot  &
    \cdot  & \cdot  & \widetilde{d}_{02} & \cdot  &
    \cdot  & \widetilde{d}_{02} & \cdot  & \cdot    \\   \hline
    %%%%%%%%%%%%%%%%%%%%%%%%%%%%%%%%%%%%%%%%%%
    \cdot  & \cdot  & \cdot  & \widetilde{d}_{10} &
    \widetilde{d}_{11} & \cdot  & \cdot  & \cdot  &
    \cdot  & \cdot  & \widetilde{d}_{12} & \cdot  &
    \cdot  & \widetilde{d}_{13} & \cdot  & \cdot    \\
    \widetilde{a}_{10} & \cdot  & \cdot  & \cdot  &
    \cdot  & \widetilde{a}_{11} & \cdot  & \cdot  &
    \cdot  & \cdot  & \cdot  & \widetilde{a}_{12} &
    \cdot  & \cdot  & \widetilde{a}_{13} & \cdot      \\
    \cdot  & \widetilde{b}_{10} & \cdot  & \cdot  &
    \cdot  & \cdot  & \widetilde{b}_{11} & \cdot  &
    \widetilde{b}_{12} & \cdot  & \cdot  & \cdot  &
    \cdot  & \cdot  & \cdot  & \widetilde{b}_{13}    \\
    \cdot  & \cdot  & \widetilde{c}_{10} & \cdot  &
    \cdot  & \cdot  & \cdot  & \widetilde{c}_{11} &
    \cdot  & \widetilde{c}_{12} & \cdot  & \cdot  &
    \widetilde{c}_{13} & \cdot  & \cdot  & \cdot     \\   \hline
    %%%%%%%%%%%%%%%%%%%%%%%%%%%%%%%%%%%%%%%%%%
    \cdot  & \widetilde{b}_{20} & \cdot  & \cdot  &
    \cdot  & \cdot  & \widetilde{b}_{21} & \cdot  &
    \widetilde{b}_{22} & \cdot  & \cdot  & \cdot  &
    \cdot  & \cdot  & \cdot  & \widetilde{b}_{23}    \\
    \cdot  & \cdot  & \widetilde{c}_{20} & \cdot  &
    \cdot  & \cdot  & \cdot  & \widetilde{c}_{21} &
    \cdot  & \widetilde{c}_{22} & \cdot  & \cdot  &
    \widetilde{c}_{23} & \cdot  & \cdot  & \cdot     \\
    \cdot  & \cdot  & \cdot  & \widetilde{d}_{20} &
    \widetilde{d}_{21} & \cdot  & \cdot  & \cdot  &
    \cdot  & \cdot  & \widetilde{d}_{22} & \cdot  &
    \cdot  & \widetilde{d}_{22} & \cdot  & \cdot    \\
    \widetilde{a}_{20} & \cdot  & \cdot  & \cdot  &
    \cdot  & \widetilde{a}_{21} & \cdot  & \cdot  &
    \cdot  & \cdot  & \cdot  & \widetilde{a}_{22} &
    \cdot  & \cdot  & \widetilde{a}_{23} & \cdot     \\   \hline
    %%%%%%%%%%%%%%%%%%%%%%%%%%%%%%%%%%%%%%%%%%
    \cdot  & \cdot  & \widetilde{c}_{30} & \cdot  &
    \cdot  & \cdot  & \cdot  & \widetilde{c}_{31} &
    \cdot  & \widetilde{c}_{32} & \cdot  & \cdot  &
    \widetilde{c}_{33} & \cdot  & \cdot  & \cdot     \\
    \cdot  & \cdot  & \cdot  & \widetilde{d}_{30} &
    \widetilde{d}_{31} & \cdot  & \cdot  & \cdot  &
    \cdot  & \cdot  & \widetilde{d}_{32} & \cdot  &
    \cdot  & \widetilde{d}_{32} & \cdot  & \cdot    \\
    \widetilde{a}_{30} & \cdot  & \cdot  & \cdot  &
    \cdot  & \widetilde{a}_{31} & \cdot  & \cdot  &
    \cdot  & \cdot  & \cdot  & \widetilde{a}_{32} &
    \cdot  & \cdot  & \widetilde{a}_{33} & \cdot      \\
    \cdot  & \widetilde{b}_{30} & \cdot  & \cdot  &
    \cdot  & \cdot  & \widetilde{b}_{31} & \cdot  &
    \widetilde{b}_{32} & \cdot  & \cdot  & \cdot  &
    \cdot  & \cdot  & \cdot  & \widetilde{b}_{33}
\end{array} \right)\ .
\end{equation}
\end{widetext}
where the matrices
$\widetilde{a},\widetilde{b},\widetilde{c},\widetilde{d}$ are
given by
\begin{widetext}
\[
%\begin{equation*}\label{}
\widetilde{a}\ = \ \left( \begin{array}{cccc}
    a_{00} & b_{01} & d_{02} & c_{03} \\
    b_{10} & c_{11} & a_{12} & d_{13}   \\
    d_{20} & a_{21} & c_{22} & b_{23} \\
    c_{30} & d_{31} & b_{32} & a_{33} \end{array} \right)\ , \ \
    \widetilde{b}\ = \ \left( \begin{array}{cccc}
    b_{00} & c_{01} & a_{02} & d_{03} \\
    c_{10} & d_{11} & b_{12} & a_{13}   \\
    a_{20} & b_{21} & d_{22} & c_{23} \\
    d_{30} & a_{31} & c_{32} & b_{33}\end{array} \right)\ , \ \
    \widetilde{c}\ = \ \left( \begin{array}{cccc}
    c_{00} & d_{01} & b_{02} & a_{03} \\
    d_{10} & a_{11} & c_{12} & b_{13}   \\
    b_{20} & c_{21} & a_{22} & d_{23} \\
    a_{30} & b_{31} & d_{32} & c_{33}\end{array} \right)\ , \ \
    \widetilde{d}\ = \ \left( \begin{array}{cccc}
    d_{00} & a_{01} & c_{02} & b_{03} \\
    a_{10} & b_{11} & d_{12} & c_{13}   \\
    c_{20} & d_{21} & b_{22} & a_{23} \\
    b_{30} & c_{31} & a_{32} & d_{33}\end{array} \right)\ .
\]
\end{widetext}


\begin{thebibliography}{1} \bibliographystyle{plain}

\bibitem{QIT} M. A. Nielsen and I. L. Chuang, {\it Quantum computation
and quantum information}, Cambridge University Press, Cambridge,
2000.

\bibitem{Horodecki-review} R. Horodecki, P. Horodecki, M. Horodecki and K.
Horodecki, {\it Quantum entanglement}, arXiv: quant-ph/0702225.


\bibitem{Peres} A. Peres, Phys. Rev. Lett. {\bf 77}, 1413 (1996).

\bibitem{PPT} P. Horodecki, Phys. Lett. A {\bf 232}, 333 (1997).

\bibitem{Horodeccy-PM} M. Horodecki, P. Horodecki and R. Horodecki, Phys. Lett. A
{\bf 223}, 1 (1996);

\bibitem{UPB} C. Bennet, D. DiVincenzo, T. Mor, P. Shor, J. Smolin and B.
Terhal, Phys. Rev. Lett. {\bf 82}, 5385 (1999).

\bibitem{UPB-inne} D.P. DiVincenco,  T. Mor, P. W. Shor, J. A. Smolin and B. M. Terhal, Comm. Math. Phys. {\bf
238}, 379 (2003); A.O. Pittenger, Lin. Alg. Appl. {\bf 359}, 235
(2003).


\bibitem{Horodecki-book}  P. Horodecki, M. Horodecki, and R. Horodecki, Phys. Rev.
Lett. {\bf 82}, 1056 (1999).

\bibitem{Doherty-PPT} A. C. Doherty, P.A. Parrilo, and F. M. Spedalieri, Phys.
Rev. Lett. {\bf 88}, 187904 (2002).

%\bibitem{DISTIL} C.H. Bennett, D.P. DiVincenzo, J.A. Smolin  and W.K. Wootters, Phys. Rev. A {\bf 54},
%3824 (1996); M. Horodecki, P. Horodecki  and R. Horodecki, Phys.
%Rev. Lett. {\bf 84},  2014 (2000).

\bibitem{Shor} D. P. DiVincenzo, P. W. Shor, J. A. Smolin, B. M. Terhal and A. V. Thapliyal, Phys.
Rev. A {\bf 61}, 062312 (2000).



\bibitem{Ex2} P. Horodecki and M. Lewenstein, Phys. Rev. Lett.
{\bf 85}, 2657 (2000).

\bibitem{Ex3} D. Bruss and A. Peres, Phys. Rev. A {\bf 61},
030301(R) (2000).

\bibitem{Ex4} S. Yu and N. Liu, Phys. Rev. Lett, {\bf 95}, 150504
(2005).

\bibitem{Piani} M. Piani and C. Mora, Phys. Rev. A {\bf 75}, 02305 (2007).


\bibitem{Clarisse} L. Clarisse,  Phys. Lett. A {\bf 359} (2006)
603.

\bibitem{Ha} K.-C. Ha, Publ. RIMS, Kyoto Univ. {\bf 34}, 591
(1998)


\bibitem{Ha1}  K. Ha, S.-H Kye, and Y. Park, Phys. Lett. A {\bf 313}, 163 (2003).

\bibitem{Ha2}  K. Ha, S.-H. Kye, Phys. Lett. A {\bf 325} 315
(2004); J. Phys. A: Math. Gen. {\bf 38}, 9039 (2005).


\bibitem{Norwegia} J. M. Leinaas, J. Myrheim and E.
Ovrum, {\it Extreme points of the set of density matrices with
positive partial transpose}, arXiv:0704.3348[quant-ph]

\bibitem{III} D. Chru\'sci\'nski and A. Kossakowski, Phys. Rev. A
{\bf 74}, 022308 (2006).

\bibitem{C} A
matrix $X \in M_n$ is circulant if
\[    X = \left( \begin{array}{cccc} \alpha_0 & \alpha_1 &
     \ldots & \alpha_{n-1} \\ \alpha_{n-1} & \alpha_0 &
     \ldots & \alpha_{n-2} \\
    \vdots &\vdots  &\ddots &\vdots \\
  \alpha_1 & \alpha_2 &
     \ldots & \alpha_0 \end{array} \right) \ . \]

\bibitem{H} A Hadamard (or Schur) product of two $n \times n$ matrices $A=[A_{ij}]$
and $B=[B_{ij}]$ is defined by
\[  (A \circ B)_{ij} = A_{ij} B_{ij}\ . \]

\bibitem{Horodecki} M. Horodecki and P. Horodecki, Phys. Rev. {\bf A} 59, 4206
(1999).

\bibitem{Werner} R.F. Werner, Phys. Rev. A {\bf 40}, 4277 (1989).

\bibitem{Robertson} A. G. Robertson, Quart. J. Math. Oxford (2),
{\bf 34}, 87 (1983).


\bibitem{EW} D. Chru\'sci\'nski and A. Kossakowski, {\it On the structure of entanglement witnesses
and new class of positive  indecomposable maps}, arXiv:
quant-ph/0606211  (to appear in Open System Inf. Dynamics).



\bibitem{Fei}  S.-M. Fei, X. Li-Jost, B.-Z. Sun, Phys. Lett. A {\bf 352}, 321
(2006).

\bibitem{edge1} M. Lewenstein, B. Kraus, J. I. Cirac and P. Horodecki,
 Phys. Rev. A {\bf 62}, 052310 (2000).

\bibitem{edge2} A. Sanpera, D. Bruss and M. Lewenstein,
Phys. Rev. A {\bf 63},  050301 (2001).

\end{thebibliography}
\end{document}